\documentclass[12pt,epsf]{article}
\usepackage{amssymb,amsmath}
\usepackage{graphicx,color}

\newcommand{\be}{\begin{equation}}
\newcommand{\ee}{\end{equation}}
\newcommand{\bea}{\begin{eqnarray}}
\newcommand{\eea}{\end{eqnarray}}
\newcommand{\beas}{\begin{eqnarray*}}
\newcommand{\eeas}{\end{eqnarray*}}
\newcommand{\ba}{\begin{array}}
\newcommand{\ea}{\end{array}}

\newcommand{\red}{\textcolor{red}}

\newcommand{\bbibitem}[1]{\bibitem{#1}\marginpar{#1}}
\def\Bibitem#1{\bibitem{#1}%
  \smash{\hbox to0pt{\raise1ex\hbox{\tiny[#1]}\hss}}}

\def\Label#1{\label{#1}%
  \smash{\hbox to0pt{\raise1ex\hbox{\tiny[#1]}\hss}}}
\def\noLabels{\let\Label=\label}
\def\nobbibitem{\let\bbibitem=\bibitem}
 \def\noBibitem{\let\Bibitem=\bibitem}

\renewcommand{\(}{\left(}
\renewcommand{\)}{\right)}
\newcommand{\tr}{\mathrm{tr}}

\def\identity{{\rlap{1} \hskip 1.6pt \hbox{1}}}
\newcommand{\nbox}{{\,\lower0.9pt\vbox{\hrule \hbox{\vrule height 0.2 cm \hskip 0.19 cm \vrule height 0.2 cm}\hrule}\,}}

\def\href#1#2{#2}

\begin{document}
\noLabels 
\noBibitem 

\begin{titlepage}
\hfill
\vbox{
    \halign{#\hfil         \cr
           } 
      }  
\vspace*{20mm}
\begin{center}
{\Large \bf Momentum-space entanglement \\ and renormalization  in quantum field theory}\\
\vspace*{15mm}
\vspace*{1mm}
Vijay Balasubramanian${}^a$, Michael B. McDermott${}^b$, Mark Van Raamsdonk${}^b$
\vspace*{1cm}

{${}^a$
{David Rittenhouse Laboratory, 209 S. 33rd Street, University of Pennsylvania, Philadelphia, PA 19104, USA} \\ ${}$ \\
${}^b$
Department of Physics and Astronomy,
University of British Columbia\\
6224 Agricultural Road,
Vancouver, B.C., V6T 1W9, Canada}

\vspace*{1cm}
\end{center}

\begin{abstract}
The degrees of freedom of any interacting quantum field theory are entangled in momentum space. Thus, in the vacuum state, the infrared degrees of freedom are described by a density matrix with an entanglement entropy.  We derive a relation between this density matrix and the conventional Wilsonian effective action. We argue that the entanglement entropy of and mutual information between subsets of field theoretic degrees of freedom at different momentum scales are natural observables in quantum field theory and demonstrate how to compute these in perturbation theory. The results may be understood heuristically based on the scale-dependence of the coupling strength and number of degrees of freedom. We measure the rate at which entanglement between degrees of freedom declines as their scales separate and suggest that this decay is related to the property of decoupling in quantum field theory.
\end{abstract}

\end{titlepage}

\vskip 1cm

\section{Introduction}

A quintessential property that distinguishes quantum mechanics from classical mechanics is the entanglement of otherwise distinct degrees of freedom. When certain degrees of freedom are entangled with the rest of a quantum system, it is not possible to describe them by a pure state. Rather, the most complete description of a subsystem $A$ at a particular time is via the reduced density matrix obtained by tracing over the degrees of freedom in the complement, $\rho_A = \tr_{\bar{A}}(|\Psi \rangle \langle \Psi |)$, where $|\Psi \rangle$ is the state of the entire system. The entropy constructed from the reduced density matrix,   $S(\rho_A) = -\tr( \rho_A \ln(\rho_A))$, quantifies the amount of entanglement between $A$ and its complement. The entanglement entropies corresponding to reduced density matrices for diverse subsets of degrees of freedom provide a rich characterization of the quantum state for systems with many degrees of freedom.\footnote{For a basic review of density matrices, entanglement entropy, and related concepts, see for example \cite{nc}.}

In physical systems, we typically only have access to a subset of the degrees of freedom, namely the low-energy or long-wavelength modes that are directly accessible to experiments. In an interacting theory, it will generally be true that these degrees of freedom are entangled with the inaccessible high-energy degrees of freedom. Thus, the long-wavelength modes will be described by a density matrix. A more familiar description of low-energy degrees of freedom is due to Wilson \cite{Wilsonrg} -- one carries out the complete path integral over the inaccessible degrees of freedom, arriving at an effective action capturing the dynamics of the remaining system. We show that the Wilsonian prescription is compatible with the description in terms of a density matrix: given a Wilsonian effective action we can canonically associate the corresponding density matrix via equation (\ref{final}) below.

For continuous physical systems described by interacting quantum field theories, understanding the variation with scale of the Wilsonian effective action $S_W(\mu)$ provides key insights into the nature of the quantum field theories, revealing a striking insensitivity of the low-energy physics to the details of the ultraviolet description. Correspondingly, it is natural to consider the variation with scale of the density matrix $\rho(\mu)$ for the degrees of freedom with momentum $|\vec{p}| < \mu$ and the associated entanglement entropy $S(\mu)$. To make our considerations concrete, we derive a formula for this low-energy entanglement entropy in perturbative quantum field theory, and apply it to scalar field theories with $\phi^n$ potentials in various dimensions. The scale-dependence of the entropy $S(\mu)$ in such theories can be understood in terms of the variation of the coupling and number of degrees of freedom with scale.

To study entanglement between scales in greater detail, we can consider the entanglement entropy associated with any subset of the allowed momenta, or the mutual information between any two subsets of the allowed momenta, for example between individual modes with momenta $\vec{p}$ and $\vec{q}$. These measures characterize the extent of entanglement between specific scales in field theory, and we compute the rate at which this entanglement declines as the scales separate. This fall-off may give an alternative characterization of the property of decoupling in quantum field theories. In theories that do not enjoy the property of decoupling, e.g. noncommutative gauge theories \cite{Minwalla:1999px} and theories of gravity, the entanglement between degrees of freedom at different scales may play an especially important role.

Entanglement entropy in quantum field theory has been considered previously, but almost all previous work has focused on entanglement between degrees of freedom associated with spatial regions (e.g. \cite{Calabrese:2004eu, Casini:2004bw} ). The notion of a density matrix for low-momentum modes or entanglement between different momentum modes has appeared earlier in the context of cosmology and condensed matter physics (e.g. \cite{Lombardo:1995fg,onemode,Mazur:2008wa}), but there is little overlap with the present work.

\section{The low-energy density matrix}
\Label{Wilson}

A quantum system with many degrees of freedom has a Hilbert space of the form ${\cal H} = {\cal H}_1 \otimes {\cal H}_2 \otimes \cdots$.   Given a subset of these degrees of freedom ($A$, with complement $\bar{A}$), we can write ${\cal H} = {\cal H}_A \otimes {\cal H}_{\bar{A}}$ where ${\cal H}_A$ is the tensor product of Hilbert spaces for the degrees of freedom in $A$.  If $\rho$ is the density matrix for the full system (which may be in a pure state), a reduced density matrix for $A$ is defined by tracing over $\bar{A}$: $\rho^A = \tr_{\bar{A}}(\rho)$ (or, given components in a specific basis, $\rho^A_{mn} = \sum_N \rho_{mN,nN}$).   Expectation values of operators that act only on $A$ can be computed as $\tr \left(\rho \; ({\cal O}_A \otimes \identity)\right) = \tr_A(\rho^A {\cal O}_A)$.  If $A$ is entangled with its complement,  $\rho^A$ will have a finite entropy: $S = -\tr_A(\rho^A \log \rho^A) > 0$.

In this construction, the Hilbert space can be be decomposed into a tensor product in any convenient way.  For example, the Hilbert space for two identical oscillators could be decomposed either as a product of the Hilbert spaces for the individual oscillators, or as a product of the Hilbert spaces of even and odd normal modes.   A reduced density matrix could be computed in either case -- good choices of decomposition are dictated by the structure of the interactions and restrictions on which degrees of freedom are accessible to measurements.

In quantum field theories, locality makes it natural to associate independent degrees of freedom with disjoint spatial domains.  Hence, given a spatial region $A$ (and complement $\bar{A}$), one can decompose the Hilbert space as ${\cal H} = {\cal H}_A \otimes {\cal H}_{\bar{A}}$ and trace over $\bar{A}$ to derive the reduced density matrix of $A$.    But since the Hamiltonians of free field theories are diagonalized by modes of fixed momentum, it is in many ways more natural to use the Fock space decomposition, ${\cal H} = \otimes_{\vec{p}} {\cal H}_{\vec{p}}$, where ${\cal H}_{\vec{p}}$ is the Hilbert space of modes of momentum $\vec{p}$.\footnote{Here, it is clearest to define the field theory as a limit of a theory on finite volume so that the tensor product is over a discrete set of allowed momenta. For a general field theory, the factors would be labeled by field type and spin/polarization in addition to momentum.} While this decomposition is motivated by considering the case of free field theory, it applies equally well once we turn on interactions, and is indeed the standard setting for computations in perturbative quantum field theory.\footnote{There is a formal sense in which turning on interactions takes one out of the original Hilbert space. However, by placing a cutoff at some energy scale much higher than any scale of interest, the Hilbert space structure of the interacting theory will be the same as that of the free theory, and a density matrix for low-momentum modes can be precisely defined. Furthermore, as we will see later, various observables related to the spectrum of the density matrix have a well-defined limit as we take the cutoff to infinity. }

In free field theory, the vacuum state is a tensor product of the Fock space vacuum states for each independent field mode --  there is no entanglement between the field modes at different momenta.  But in an interacting theory, the full vacuum state will be a superposition of Fock basis states -- hence the modes of different momenta will generally be entangled.  In this case, the reduced density matrix corresponding to a subset of the degrees of freedom ($A$) will necessarily have a finite entropy, indicating that $A$ is effectively in a mixed state if the rest of system is traced over. Now, one is most often interested in the physics of the ``infrared'' degrees of freedom that are accessible to experiment, i.e., the degrees of freedom with momenta below some scale $\mu$.   The present discussion shows that tracing over the ultraviolet, i.e. degrees of freedom with momenta above $\mu$, should lead to an infrared effective description in terms of a {\it low-energy density matrix} corresponding to a mixed state with finite entropy.

\subsubsection*{Relation between low-energy density matrix and low-energy effective action}

The standard way of studying the low-energy theory is through the Wilsonian effective action.  How is this quantity related to the low-energy density matrix? To begin, consider a bare action $S_\Lambda$ defined with a cutoff $|p| \le \Lambda$. Associated to this, we have a Hamiltonian $H_\Lambda$, which will have some ground state $|\Psi^0_\Lambda \rangle$ and corresponding density matrix $\rho^0_\Lambda = |\Psi^0_\Lambda \rangle \langle \Psi^0_\Lambda|$. This density matrix can be written as a Euclidean path integral by taking the $T \to 0$ limit of the finite temperature density matrix
 \be
\langle \hat{\phi}_y | \rho_\Lambda^T | \phi_y \rangle = {1 \over Z} \langle \hat{\phi}_y  | e^{-\beta H_\Lambda} | \phi_y \rangle
= {1 \over Z} \int_{\phi(\tau =-\beta/2) =  \phi_y }^{\phi(\tau = \beta/2) = \hat{\phi}_y } {\cal D}\phi(\tau) \, e^{-S_{\Lambda}^E} \; ,
\label{patha}
\ee
where $\{ \phi_y \}$ is a basis of field configurations indexed by $y$, $\beta = 1/T$ is the inverse temperature, $S_\Lambda^E$ is the Euclidean action, and $Z$ is the partition sum that normalizes the path integral.

Given a subset of degrees of freedom $A$ (with complement $\bar{A}$) and the tensor product structure of the Hilbert space, we can split the parameter $y$ which indexes the basis states as $y = (a, {\bar{a}})$, and a pick a basis with $\phi_y = \phi_{a} \times \phi_{\bar{a}}$.   To define a reduced density matrix for $A$ by tracing over $\bar{A}$ we write
\be
\langle \hat{\phi}_{a}|\rho_A^T | \phi_{a} \rangle = \int {\cal D} \phi'_{\bar{a}} \,    \langle \hat{\phi}_{a} \phi'_{\bar{a}} | \rho_\Lambda^T | {\phi}_{a} \phi'_{\bar{a}}\rangle
= {1 \over Z} \int_{\phi_A(-\beta/2) = \phi_a}^{\phi_A(\beta/2) = \hat{\phi}_a}  {\cal D}\phi_A(\tau) \, {\cal D}\phi_{\bar{A}}(\tau) e^{-S_{E}}\; .
\ee
In the last expression, the fields $\phi_{\bar{A}}$ are periodic in the range $[-\beta/2,\beta/2]$, which is implied after substituting (\ref{patha}) into the trace in the middle expression.

Now define a conventional  thermal effective action for the the subsystem $A$:
\be
e^{-S^T_{W}(\phi_A)} = \int_{-\beta/2 \le \tau \le \beta/2}  {\cal D}\phi_{\bar{A}}(\tau) \,  e^{-S_{E}(\phi_A,\phi_{\bar{A}})} \, .
\ee
In terms of this, the density matrix for $A$ is
\be
\langle \hat{\phi}_a |\rho_A^{T} | \phi_a \rangle = {1 \over Z} \int_{\phi_A(-\beta/2) = \phi_a}^{\phi_A( \beta/2) = \hat{\phi}_a} {\cal D}\phi_A(\tau) \,  e^{-S^T_{W}(\phi_A)}
=
{1 \over Z}\int_{\phi_A(\tau = 0^-) = \phi_a}^{\phi_A(\tau = 0^+) = \hat{\phi}_a}  {\cal D}\phi_A(\tau) \,  e^{-S^T_{W}(\phi_A)} \; .
\ee
In the last expression we translated time to put the discontinuity in the integral at $\tau = 0^\pm$ and the fields are taken to be periodic at $\tau = \pm \beta/2$. The reduced density matrix for $A$ in the ground state $|\Psi^0_\Lambda \rangle$ for the entire system is extracted by taking $\beta \to \infty$.

We now specialize to the case where $A$ is the subset of degrees of freedom with spatial momenta $|p|< \mu$ for any scale $\mu$ which is lower than the ultraviolet cutoff $\Lambda$. The reduced density matrix $\rho_{|p|<\mu}$ obtained by tracing over the degrees of freedom with momenta in the range $\mu< |p| \le \Lambda$ is thus given by:
\bea
\langle \hat{\phi}_{|p|<\mu} |\rho_{|p|<\mu} |\tilde{\phi}_{|p|<\mu} \rangle &=& {1 \over Z} \int^{\phi_{|p|<\mu}(\tau = 0^+) = \hat{\phi}_{|p|<\mu}}_{\phi_{|p|<\mu}(\tau = 0^-) = \tilde{\phi}_{|p|<\mu}} {\cal D}\phi_{|p|<\mu}  \,e^{-S_{W}(\phi_{|p|<\mu})}\; .
\Label{final}
\eea
where now (having taken $\beta \to \infty$,)  $S_{W}$ is the standard Wilsonian effective action obtained by integrating out the degrees of freedom with large spatial momenta:\footnote{Note that while the path integral is Euclidean, we are integrating out all modes with spatial momenta $|p|>\mu$, regardless of frequency.}
\be
e^{-S_{W}(\phi_{|p|<\mu})} = \int  {\cal D} \phi_{|p|>\mu}(\tau) \,  e^{-S_{E}(\phi_{|p|<\mu},\phi_{|p|>\mu})} \, \; .
\ee

Equation (\ref{final}) is our final result for the low-energy density matrix. In particular, if ${\cal O}$ is an observable built out of the low-momentum modes at $\tau=0$, it follows from (\ref{final}) that
\be
\tr(\hat{{\cal O}} \rho) =  {1 \over Z} \int [d \phi_{|p|<\mu}] {\cal O} e^{-S_{W}(\phi_{|p|<\mu})} \; ,
\ee
so the standard calculation using the effective action is equivalent to a calculation using the density matrix. Of course, the full Wilsonian effective action contains more information than the density matrix associated with the vacuum state of the field theory. The former is a functional of time-dependent field configurations, while the latter depends only on a pair of {\it time-independent} field configurations.

The description of low-energy degrees of freedom via a density matrix may seem unfamiliar and one may ask why we cannot simply associate a pure vacuum state to the low-energy degrees of freedom based on the effective action. The reason is that $S_{W}$ will typically contain terms with higher time derivatives, there is no way to associate to $S_W$ a Hamiltonian $H_\mu$ expressed exclusively in terms of the low-momentum variables and their conjugate momenta. Thus, there is no canonical way to associate a pure state of the low-momentum part of the Hilbert space to the full ground state of the theory. As we have seen, the object that can be canonically associated to a Wilsonian effective action for these low-momentum degrees of freedom is a density matrix.

\section{Measures of entanglement}
\Label{entanglement}

What observables quantify the amount of entanglement between the degrees of freedom in different ranges of momenta?   In this section we begin by discussing such quantities in generality and conclude by constructing perturbative expressions for such observables in weakly coupled field theories.

First,  for any density matrix $\rho$, the von Neumann entropy
\be
S(\rho) = -\tr(\rho \ln(\rho))
\ee
measures the classical uncertainty associated with the mixed state described by $\rho$.  When $\rho$ represents a microcanonical or canonical ensemble, the von Neumann entropy gives the thermodynamic entropy.   When $\rho$ is the reduced density matrix describing a subsystem $A$ of a quantum system that is in a pure state, $S$ quantifies the  entanglement between $A$ and its complement ($\bar{A}$). In this case the {\it entanglement entropy}  of $A$ is equal to that of $\bar{A}$, a fact that follows from a stronger result that the spectrum of eigenvalues of $\rho_A$ matches the spectrum of eigenvalues of $\rho_{\bar{A}}$.

When the Hilbert space for the theory can be decomposed into a tensor product with three or more factors, the quantum entanglement and classical correlations between pairs of these subsystems are jointly quantified by the {\it mutual information}.  For instance, if the Hilbert space is of the form ${\cal H} = {\cal H}_A \otimes {\cal H}_B \otimes {\cal H}_C \otimes \cdots $,
the {\it mutual information} between $A$ and $B$ is defined as
\be
I(A,B) = S(A) + S(B) - S(A \cup B) \; .
\ee
where $S(X)$ is the von Neumann entropy of the reduced density matrix of subsystem $X$. Mutual information is always greater than or equal to zero, with equality if and only if the density matrix for the $AB$ subsystem is a tensor product of the reduced density matrices for subsystems $A$ and $B$. In other words, mutual information is zero if and only if there is neither any entanglement nor any classical correlation between the two subsystems.\footnote{The systems $A$ and $B$ are said to be entangled if and only if the density matrix for the $AB$ subsystem cannot be written as $\sum_i p_i \rho_A^i \otimes \rho_B^i$.  {\it Separable} density matrices of this form represent states which have no quantum entanglement, but may have classical correlations. The mutual information for such a state can be nonzero.} Mutual information provides an upper bound on all correlators between the two regions:
for any bounded operators ${\cal O}_A$ and ${\cal O}_B$, acting only on the subsystems $A$ and $B$, we have \cite{wvhc}
\be
\label{corrbound}
I(A,B) \ge \frac{(\langle {\cal O}_A {\cal O}_B \rangle - \langle {\cal O}_A \rangle \langle {\cal O}_B \rangle  )^2}{2|{\cal O}_A|^2 |{\cal O}_B|^2} \; .
\ee

If the Hilbert space consists of three factors  ${\cal H} = {\cal H}_A \otimes {\cal H}_B \otimes {\cal H}_C$ and the complete system is in a pure state it follows from the definitions that
\be
I(A \cup B,C) = I(A,C) + I(B,C) \; .
\ee
But  if $A$, $B$, and $C$ together comprise only a part of the system, another interesting observable is the {\it tripartite information} which quantifies the extent to which the mutual information between $A \cup B$ and  $C$ is determined by the parirwise mutual informations $I(A,B)$ and $I(B,C)$:
\be
I(A,B,C) = I(A \cup B,C) - I(A,C) - I(B,C)  \, .
\ee
In general, this quantity can be positive, negative, or zero. For a pure state of the full system, $I(A,B,C)$ is symmetric between the four subsystems $A$,$B$,$C$, and $\overline{A \cup B \cup C}$.

\subsection{Entanglement observables in perturbation theory}
\Label{sec:entpert}

For weakly coupled quantum field theories, we can use perturbation theory methods to calculate the entanglement observables described in the previous section. To begin, it is useful to derive a set of perturbative results that apply more broadly.

Consider a general quantum system whose Hilbert space may be decomposed into a tensor product ${\cal H} = {\cal H}_A \otimes {\cal H}_B$, and start with a Hamiltonian of the form
\be
H = H_A \otimes \identity + \identity \otimes H_B \; .
\ee
Denote the energy eigenstates of $H_A$ by $|n \rangle$ and the energy eigenstates of $H_B$ by $|N \rangle$, with energies $E_n$ and $\tilde{E}_N$ respectively. Before adding interactions, the ground state is
\be
|0 ,0 \rangle  \equiv |0 \rangle \otimes |0 \rangle \; .
\ee
Now,  perturb the Hamiltonian by an interaction $\lambda H_{AB}$, where $\lambda$ is  a small parameter. The perturbed ground state may be written (before normalization) as
\be
|\Omega \rangle = |0 ,0 \rangle + \sum_{n \ne 0} A_n |n ,0 \rangle + \sum_{N \ne 0} B_N |0,N \rangle + \sum_{n, N \ne 0} C_{n,N} |n,N \rangle \, ,
\Label{pertstate}
\ee
where $A$,$B$, and $C$ are coefficients starting at order $\lambda$ that may be computed in perturbation theory. To normalize, we should multiply by $1/{\cal N}^{1 \over 2}$, where
$
{\cal N} = 1 + \sum |A_{n}|^2 + \sum |B_{N}|^2  + \sum |C_{n,N}|^2 \; .
$
Now, the density matrix corresponding to the subsystem $A$ is
\be
\rho_A = {1 \over 1 + |A|^2 + |B|^2 + |C|^2} \left( \ba{cc} 1 + |B|^2 & A^\dagger + B C^\dagger \cr A + C B^\dagger & A A^\dagger + C C^\dagger \ea \right) \; ,
\Label{densitymat1}
\ee
where the elements of this matrix correspond to $|0 \rangle \langle 0 |$,$|0 \rangle \langle n|$,$|m \rangle \langle 0 |$,$|m \rangle \langle n |$ terms respectively. By a symmetry transformation $\rho \to M \rho M^{-1}$, we can simplify the form to
\be
\hat{\rho}_A = \left( \ba{cc} 1 - |C|^2  & 0 \cr 0 & C C^\dagger \ea \right) + {\cal O}(\lambda^3)
\Label{densitysimp}
\ee
where we are using the fact that $A$, $B$, and $C$ start at order $\lambda$.  (See below for why this is possible.)

Up to corrections of order $\lambda^3$, the eigenvalues of this matrix are $\lambda^2 a_i$ and $1 - \lambda^2 \sum a_i$, where $\{a_i \}$  (normalized to be of order $\lambda^0$) are the eigenvalues of the matrix $C C^\dagger / \lambda^2$.   Thus, the entanglement entropy is
\bea
\label{evexp}
S_A = - \tr(\rho_A \log(\rho_A)) &= - (1 - \lambda^2 \sum a_i) \log (1 - \lambda^2 \sum a_i)  - \sum \lambda^2 a_i \log(\lambda^2 a_i) \cr
&= - \lambda^2 \log(\lambda^2) \sum a_i + \lambda^2 \sum a_i (1 - \log a_i) + {\cal O}(\lambda^3) \; .
\eea
Now, an explicit expression for the $C$ matrix  using standard perturbation theory is
\be
C_{nN} = \lambda { \langle n,N | H_{AB} | 0, 0 \rangle \over E_0 + \tilde{E}_0 - E_n - \tilde{E}_N} + {\cal O} (\lambda^2) \; .
\ee
Using this, the leading term in the entanglement entropy for small $\lambda$ is explicitly
\be
\label{master}
S_A = - \lambda^2 \log(\lambda^2) \sum_{n \ne 0, N \ne 0} { |\langle n,N | H_{AB} | 0, 0 \rangle|^2  \over (E_0 + \tilde{E}_0 - E_n - \tilde{E}_N)^2} + {\cal O} (\lambda^2) \, .
\ee
Interestingly, the entanglement entropy is not analytic in $\lambda$ at $\lambda = 0$.   Also, the leading order perturbative result (up to order $\lambda^2$ terms which are not written explicitly) depends only on matrix elements of the interaction Hamiltonian between the vacuum and states with both subsystems excited. This follows since
(\ref{pertstate}) can be written as
\be
|\Omega \rangle = (|0 \rangle + \sum_{n \ne 0} A_n |n \rangle) \otimes (|0 \rangle + \sum_{N \ne 0} B_N |0,N \rangle) + \sum_{n, N \ne 0} (C_{n,N} - A_n B_N)|n\rangle \otimes |N \rangle \; .
\ee
In this expression, the entanglement would be zero without the second term, and in this term, $C_{nN}$ starts at order $\lambda$ while $A_n B_N$ starts at order $\lambda^2$.  The $A$ and $B$ coefficients do appear in the order $\lambda^3$ contributions to the entanglement entropy (see Appendix~\ref{app:higher}).

\subsubsection*{Mutual information}

By a similar calculation, starting from a pure state in a theory with ${\cal H} = {\cal H}_A \otimes {\cal H}_B \otimes {\cal H}_C$,
the leading contribution to $I(A,B)$ in perturbation theory is
\be
\label{master2}
I(A,B) = - \lambda^2 \log(\lambda^2)\left\{ 2 \sum_{N_A \ne 0, N_B \ne 0, N_C = 0} + \sum_{N_A \ne 0, N_B \ne 0, N_C \ne 0} \right\} { |\langle N_A, N_B,N_C | H_{int} | 0, 0 \rangle|^2  \over (E_0  - E_{N_i})^2}
\ee
Similarly, when ${\cal H} = {\cal H}_A \otimes {\cal H}_B \otimes {\cal H}_C \otimes {\cal H}_D$,
at leading order in perturbation theory, the tripartite information $I(A,B,C)$ is
\be
\label{master3}
I(A,B,C) = + \lambda^2 \log(\lambda^2) \sum_{N_i \ne 0} { |\langle N_A, N_B,N_C,N_D | H_{int} | 0, 0 \rangle|^2  \over (E_0  - E_{N_i})^2} + {\cal O} (\lambda^2)
\ee
While $I(A,B,C)$ can in general be positive, negative or zero, we see that the leading perturbative result for $I(A,B,C)$ is always less than or equal to zero. This implies that to leading order in perturbation theory,
\be
I(A \cup B,C) \le I(A,C) + I(B,C) \; .
\ee
This result is not true for general systems.\footnote{In particular, if $A$ and $B$ are completely uncorrelated, $\rho_{AB} = \rho_A \otimes \rho_B$, the opposite inequality, $I(A \cup B,C) \ge I(A,C) + I(B,C)$ follows from strong subadditivity of entanglement entropy. } Note also that if the matrix element of the interaction Hamiltonian between the free vacuum and states with all four subsystems excited is zero\footnote{In field theory, this will be true for theories with only cubic interaction terms.} then we will have
\be
\label{miequal}
I(A \cup B,C) = I(A,C) + I(B,C) \; .
\ee
to leading order in perturbation theory. In this case, the leading order contribution to mutual information between any two subsystems is completely determined from the mutual information between any pair of minimal subsystems.\footnote{In field theory, such minimal subsystems will be the Hilbert spaces associated with modes at a single momentum.}

\subsection{Entanglement observables in quantum field theory}

For all of the observables discussed above, the essential quantities we have to compute are the reduced density matrices of the various subsystems.   Given these quantities, we can compute the associated von Neumann entropies and mutual informations.  In local quantum field theory, recent discussions of entanglement have focused on the density matrices associated with bounded spatial regions. These are well-defined because (by locality) there are independent degrees of freedom in disjoint spatial domains, so the Hilbert space factorizes as required. The associated spatial entanglement entropy is typically divergent, even in free field theory, because in the continuum limit any spatial region contains an infinite number of degrees of freedom at arbitrarily short wavelengths. These divergences require regularization (e.g. by including an ultraviolet cutoff) and some care is needed to extract finite regularization-independent data \cite{Casini:2004bw}.

Now, as discussed above, it is often more natural in quantum field theory to organize degrees of freedom by momentum (or wavelength).   Corresponding to any bounded subset of momenta in a field theory there are a finite number of degrees of freedom per unit spatial volume.\footnote{For a field theory at finite volume there will be a finite number of degrees of freedom in a bounded range of momenta.   In the infinite volume limit, the set of allowed momenta becomes continuous. While there are now an infinite number of degrees of freedom with momenta in a finite region of momentum space, the number per unit spatial volume remains finite.}  As a result, the entanglement entropy associated with such a subset should be {\it finite} for a finite volume system, increasing with the volume considered.  For a translation-invariant state, we expect that the momentum space entanglement entropy will be an extensive quantity with a finite {\it density} $S/V$. We will verify this below.

What observables can we compute? We can define the entanglement entropy $S(P)$ associated to any subset $P$ of the allowed momenta\footnote{More generally, $P$ could represent a subset of the allowed single particle states.}, the mutual information $I(P,Q)$ between any two subsets of momenta, or the tripartite entanglement $I(P,Q,R)$ for any three subsets of momenta. We will focus on
\begin{itemize}
\item
$S(\mu)$, the entanglement entropy between all degrees of freedom with momenta above and below the scale $\mu$.
\item
$S([\mu_1,\mu_2])$, the entanglement entropy for a shell of momenta $\mu_1 \le |p| \le \mu_2$.
\item
$S(\vec{p})$, the entanglement entropy for a single mode with momentum $p$.
\item
$I(\{|p|<\mu_1\}, \{|p|>\mu_2 \})$, the mutual information between degrees of freedom with momenta below a scale $\mu_1$ and degrees of freedom above the scale $\mu_2$.
\item
$I(\vec{p}, \vec{q})$, the mutual information between modes with momenta $\vec{p}$ and $\vec{q}$.
\end{itemize}
These quantities probe the strength and extent of entanglement in momentum space.

In free field theory, the Hamiltonian does not couple degrees of freedom with different momenta  and thus all these measures of entanglement in momentum space should vanish. Adding a weak interaction term that couples degrees of freedom with different momenta modifies the ground state and should introduce a small amount of entanglement between the various field modes.    We can characterize this entanglement  in perturbative quantum field theory by adapting the general results derived above. For the calculation of entanglement entropy, the two subsystems correspond to two complementary subsets $A$ and $\bar{A}$ of the allowed momenta. The eigenstates $|n, N \rangle$ of the unperturbed Hamiltonian are elements of the Fock space basis $|\{n_i\},\{N_I\} \rangle$, where $n_i$ and $N_I$ are occupation numbers for particle states in the two subsets. The interaction Hamiltonian takes the form
\[
H_I = \int d^d x \, {\cal H}_I(x)
\]
for some local Hamiltonian density ${\cal H}_I(x)$ that is polynomial in the fields and their derivatives. The matrix elements
\be
\label{mel}
\langle \{n_i\},\{N_I\} | H_I | 0,0 \rangle
\ee
may be computed by expanding the interaction Hamiltonian density in terms of creation and annihilation operators. The sum in (\ref{master}) is now over all states with at least one particle having momentum in the subset $A$ and at least one particle having momentum in the subset $\bar{A}$. The nonzero matrix elements (\ref{mel}) in the sum involve states with at most $k$ momenta, where $k$ is the number of fields in the interaction, and the momenta must add to zero since translation-invariance of the interaction Hamiltonian leads to a momentum-conserving delta function.

The leading contribution (\ref{master}) to the entanglement entropy can be rewritten in terms of a projected two-point correlator of the interaction Hamiltonian (see Appendix).  Below we will work directly with the expression (\ref{master}).

\section{Scalar field theory: entanglement between scales}
\Label{scalar}

To develop some concrete examples of the general formalism, we will calculate  momentum space entanglement entropy in $d+1$ dimensional scalar theories with action:
\be
S = \int d^{d+1} x ({1 \over 2} (\partial_\mu \phi)^2 - {1 \over 2} m^2 \phi^2 - {\lambda \over n!} \phi^n) \; .
\ee
For ease of formulation, we will first take the theory to be defined in a box of size $L$ with periodic boundary conditions, and assume a UV cutoff at a scale $\Lambda$.


We will compute the entanglement entropy $S(\mu)$ of degrees of freedom with momenta $|p| < \mu$ with the high-momentum modes. Denoting by $p_i$ and $P_i$ the allowed momenta below and above $\mu$,  the Fock space basis elements are written as $|\{n_{p_i} \} \rangle \otimes |\{n_{P_i} \} \rangle$.
To use the general formula (\ref{master}) for the leading contribution to the entanglement entropy, we need matrix elements of the interaction Hamiltonian between the free vacuum and the states with both low and high momenta excited.  Recall that the fields can be expanded in terms of creation and annihilation operators as
\be
\phi(x) = {1 \over L^{d \over 2}} \sum_p {1 \over \sqrt{2 \omega_p}} (a_p e^{-i p \cdot x} + a^\dagger_p e^{i p \cdot x})  \;
\ee
where $\omega_p = \sqrt{p^2 + m^2}$. The nonzero matrix elements for the interaction Hamiltonian between the Fock space vacuum and states with $n$ particles excited\footnote{We are only interested in matrix elements between the vacuum and states with at least one low-momentum particle and at least one high-momentum particle. For $\phi^3$ and $\phi^4$ field theory, the only such non-zero matrix elements have 3 and 4 particles excited respectively. For $\phi^n$ theory with $n>4$, matrix elements with $n-2k \ge 3$ particle states can also contribute, but for these theories we must also include $\phi^{n-2k}$ counterterms in the action. For now, we focus on the case of $\phi^3$ and $\phi^4$ theory.} are
\be
\langle p_1 \cdots  p_n| H_I | 0 \rangle = {1 \over 2^{n \over 2} L^{d({n \over 2}-1)}} {\delta_{p_1 + \dots + p_n} \over \sqrt{\omega_1 \cdots \omega_n}}
\ee
From (\ref{master}), we then have
\be
\label{result}
S(\mu) =  - \lambda^2 \log(\lambda^2) \sum_{ \{p_i\}_\mu} {\delta_{p_1 + \dots + p_n} \over 2^n L^{d(n-2)} \omega_1 \cdots \omega_n (\omega_1 + \dots + \omega_n)^2 } + {\cal O} (\lambda^2)
\ee
where the sum is over distinct sets of spatial momenta such that at least one momentum is below the scale $\mu$ and at least one momentum is above the scale $\mu$. More generally, the entanglement entropy for the field modes with momenta in some set $A$ is given by the same formula with the sum over distinct sets of momenta such that at least one momentum is in a set $A$ and at least one momentum is in $\bar{A}$.


It is straightforward to take the limit of infinite volume. By the usual replacements
\[
\sum_{\vec{p}} \to \left({L \over 2 \pi}\right)^d \int d^d p \qquad \qquad \delta_{\vec{p}} \to \left( {2 \pi \over L} \right)^d \delta^d(p)
\]
we find that the entanglement entropy density $S(\mu)/V$ has a well-defined limit:
\be
\label{result2}
S(\mu)/L^d =  - \lambda^2 \log(\lambda^2) { 1 \over (2 \pi)^{d(n-1)} 2^n} \int_{ \{p_i\}_\mu} \prod d^d p_i {\delta(p_1 + \dots + p_n) \over  \omega_1 \cdots \omega_n (\omega_1 + \dots + \omega_n)^2 } + {\cal O} (\lambda^2) \; .
\ee
Here, the integral is again over distinct sets of momenta such that at least one momentum is below the scale $\mu$ and at least one momentum is above the scale $\mu$.

In practice, it is often simplest to calculate the derivative $dS/d\mu$, since the $\mu$-dependence comes only in the domain of integration, and this domain for $S(\mu + d \mu)$ is almost the same as for $S(\mu)$. In the difference
\[
S(\mu + d \mu) - S(\mu)
\]
the only contributions that don't cancel between the two terms are a positive contribution in which one momentum is in the range $[\mu, \mu + d \mu]$ and all the other momenta have magnitude larger than $\mu$, and a negative contribution in which one momentum is in the range $[\mu, \mu + d \mu]$ and all the other momenta have magnitude smaller than $\mu$.

\subsection{The $\phi^3$ theory in 1+1 dimensions}

The simplest example is the $\phi^3$ theory in 1+1 dimensions.\footnote{We work with a massive $\phi^3$ theory, so the theory is perturbatively stable. We can assume higher order interaction terms $\phi^{2n}$ which stabilize the theory non-perturbatively but do not affect our leading-order perturbative calculations.} From (\ref{result2}),
\beas
S(\mu)/V &=&  - \lambda^2 \log(\lambda^2) { 1 \over 32 \pi^{2}} \int_{ \{p_i\}_\mu} \prod d p_i {\delta(p_1 + p_2 + p_3) \over  \omega_1 \omega_2 \omega_3 (\omega_1 + \omega_2 + \omega_3)^2 } + {\cal O} (\lambda^2) \cr
&\equiv& - \lambda^2 \log(\lambda^2) { 1 \over 32 \pi^{2}} I(\mu)\; .
\eeas
Letting
\be
\label{defJ}
J(p_1,p_2,p_3) = {1 \over \omega_1 \omega_2 \omega_3 (\omega_1 + \omega_2 + \omega_3)^2}
\ee
we find that
\be
{1 \over 2} {dI \over d \mu} = \int_\mu^\infty dp \, J(\mu,p,-p-\mu) - \int_{-\mu/2}^0 dp \, J(\mu,p,-p-\mu) \; .
\ee
Evaluating the right hand side analytically for large and small $\mu$, we find that\footnote{To find $I(\mu)$ we evaluate the expression for $dI/d\mu$ and then integrate with respect to $\mu$.  The constant of integration is fixed by requiring that the entanglement entropy vanish as $\mu \to 0$.}
\be
I(\mu) \to \left\{ \ba{ll} {\mu \over m^4} (\pi - {8 \sqrt{3} \over 27} \pi - {4 \over 3}) & \mu \ll m \cr
{1 \over 12 \mu^3}\left\{ {23 \over 12} + \ln \left( {\mu^2 \over m^2} \right) \right\} & \mu \gg m \ea \right.
\ee
As discussed further below, the linear behaviour for small $\mu$ is related to the linear growth in the number of degrees of freedom below scale $\mu$, while the falloff at large $\mu$ is related to fact that a $\phi^3$ is relevant in 1+1 dimensions so that the physics at large scales approaches that of the free theory, for which there is no entanglement between modes at different momenta.

\subsubsection*{Order $\lambda^2$ terms}
Above we computed the $O(\lambda^2 \log(\lambda^2))$ term in the entanglement entropy that dominates at infinitesimal $\lambda$.  At small, but finite $\lambda$, the $O(\lambda^2)$ term in (\ref{evexp}) could compete with this.   To calculate this term we must determine the eigenvalues (and not just the trace) of the matrix $C C^\dagger/\lambda^2$, where
\be
C_{\{p_{i} \}, \{P_{i}\}} = - {\lambda \over L^{1 \over 2} 2^{3 \over 2}} {\delta_{\sum p_i + \sum P_i} \over ( \sum \omega_{p_i}+ \sum \omega_{P_i} ) \sqrt{\prod \omega_{p_i} \prod \omega_{P_i}}}
\ee
and the sets $\{p_{i} \}$ and $\{P_{i}\}$ must have either one and two elements or two and one elements. Thus the matrix $M = C C^\dagger/\lambda^2$ has nonzero elements of the form $M_{p,q}$ and $M_{\{p_1,p_2 \},\{q_1,q_2\}}$. We have
\be
M_{p,q} = \delta_{p,q} {1 \over 8 L} \sum_{P,Q} {\delta_{p + P + Q} \over \omega_p \omega_Q \omega_P (\omega_p+ \omega_Q+\omega_P)^2} \; .
\ee
Thus, for each $p$ with $|p|<\mu$ we have one eigenvalue
\be
a_p = {1 \over 8 L} \sum_{|P| > \mu,|Q| > \mu} {\delta_{p + P + Q} \over \omega_p \omega_P \omega_Q (\omega_p+ \omega_P +\omega_Q)^2} \; .
\ee
The remaining block of the matrix $M$ has entries
\[
M_{\{p_1,p_2 \},\{q_1,q_2\}} = {\delta^*_{p_1 + p_2, q_1 + q_2}  \over 8 L} {\over \sqrt{ \omega_{p_1} \omega_{p_2} \omega_{q_1} \omega_{q_2}} \omega_{p_1 + p_2} ( \omega_{p_1} + \omega_{p_2}+ \omega_{p_1 + p_2})( \omega_{q_1} + \omega_{q_2}+ \omega_{p_1 + p_2})} \;
\]
where $\delta^*$ indicates that we must have $|p_1 + p_2| > \mu$ for a nonzero result. To find the remaining eigenvalues, we put $M$  in block diagonal form, with one block for each $P$ with $|P|>\mu$, where $p_1 + p_2 = q_1 + q_2 = P$. For the block labeled by $P$, we can label the matrix entries by $p_1$ and $q_1$, with
\[
M_{p_1,q_1} = {1 \over 8 L} {1 \over \sqrt{ \omega_{p_1} \omega_{P-p_1} \omega_{q_1} \omega_{P-q_1}} \omega_{P} ( \omega_{p_1} + \omega_{P-p_1}+ \omega_{P})( \omega_{q_1} + \omega_{P-q_1}+ \omega_{P})} =  { V(p_1) V(q_1) \over 8L}
\]
where
\be
V(p) = {1 \over \sqrt{ \omega_{p} \omega_{P-p} \omega_{P}} ( \omega_{p} + \omega_{P-p}+ \omega_{P})} \; .
\ee
A matrix of this form has only one nonzero eigenvalue, equal to
\be
a_P = {1 \over 8 L} \sum_{|p| < \mu , |q| < \mu} {\delta_{p + q + P} \over \omega_p \omega_q \omega_P (\omega_p+ \omega_q +\omega_P)^2} \; .
\ee
We have have one such eigenvalue for each $P$ with $|P| > \mu$.

Having found all the eigenvalues of $C C^\dagger/\lambda^2$, we can  use (\ref{evexp}) to write an expression for $S(\mu)$ including the order $\lambda^2$ term.   Recall that $S(\mu) = \lambda^2 ( 1 - \log \left(  \lambda^2 \right)) \sum a_i - \lambda^2 \sum a_i \log(a_i)$.  Taking  $L \to \infty$,
\be
\sum a_i/L = \int {d p_1 \over 2 \pi} I(p_1)
\Label{nolog}
\ee
and
\be
\sum a_i \log(a_i) /L = \int {d p_1 \over 2 \pi} I(p_1) \log(I(p_1))
\Label{log}
\ee
where
\be
I(p_1) = \int_* {dp_2 dp_3 \over 16 \pi} {\delta(p_1 + p_2 + p_3) \over \omega_1 \omega_2 \omega_3 (\omega_1 + \omega_2 + \omega_3)^2} \; .
\ee
Here, the asterisk indicates that $p_2 < p_3$, and that $p_2$ and $p_3$ must have magnitude less than $\mu$ if $p_1$ has magnitude greater than $\mu$, while $p_2$ and $p_3$ must have magnitude greater than $\mu$ if $p_1$ has magnitude less than $\mu$.

\begin{figure}
\centering
\includegraphics[width=0.5\textwidth,angle=270]{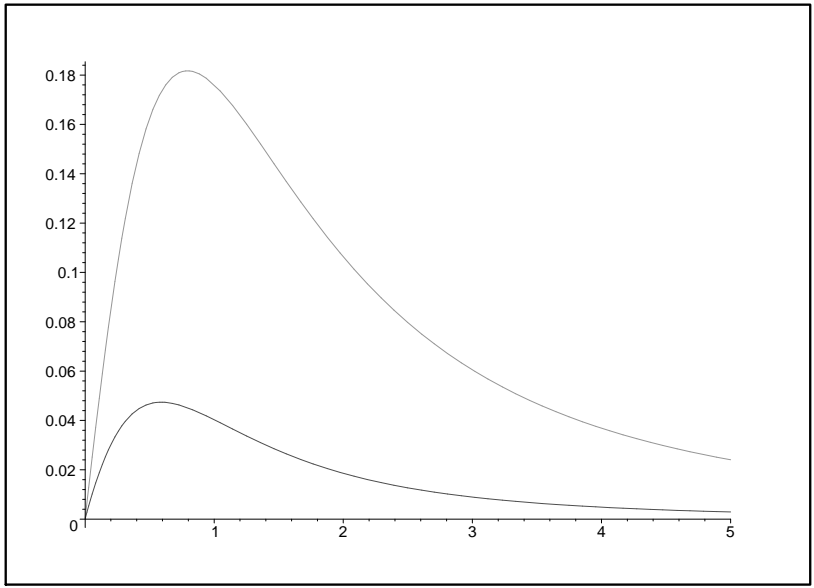}
\caption{Leading contributions to $S(\mu)$ for $\phi^3$ theory in 1+1 dimensions. Full result for $S(\mu)$ is proportional to $\lambda^2 ( \log ( 1/ \lambda^2) + 1)$ times bottom function plus $\lambda^2$ times top function.}
\label{lognolog}
\end{figure}

We have plotted the two leading contributions (\ref{log}) and (\ref{nolog}) in figure \ref{lognolog}. We see that the two terms have a qualitatively similar behavior. In detail, the term (\ref{log}) falls off slightly more slowly for large $\mu$, behaving as $1/\mu^3 \log^2(\mu^2/m^2)$ compared with $1/\mu^3 \log(\mu^2/m^2)$ for (\ref{nolog}). Thus, for fixed $\lambda$ and sufficiently large $\mu$ (of order $m/\lambda$), the ${\cal O}(\lambda^2)$ term will be larger than the ${\cal O}(\lambda^2 \log(\lambda^2))$ term, although the qualitative behavior is similar. In this work, our focus is on the physics in the limit of small $\lambda$, so in the remainder of the discussion we will concentrate ${\cal O}(\lambda^2 \log(\lambda^2))$ terms which dominate as long as we stay below the parametrically large energies of order $1/ \lambda$ relative to the mass.

\subsection{The $\phi^3$ theory in higher dimensions}

In general dimensions, the entanglement entropy for the modes below scale $\mu$ in the $\phi^3$ field theory is given by
\bea
S(\mu)/L^d &=&  - \lambda^2 \log(\lambda^2) { 1 \over 8 (2 \pi)^{2d}} \int_{ \{p_i\}_\mu}  d^d p_1 d^d p_2 d^d p_3 {\delta(p_1 + p_2 + p_3) \over  \omega_1 \omega_2 \omega_3 (\omega_1 + \omega_2 + \omega_3)^2 } + {\cal O} (\lambda^2) \cr
&\equiv& - \lambda^2 \log(\lambda^2) { 1 \over 8 (2 \pi)^{2d}} I_d(\mu)\; .
\Label{leading3}
\eea
It is more convenient to compute
\be
{1 \over \omega_{d-1} \mu^{d-1}} {dI_d \over d \mu} = (\int_{B} - \int_A )d^2 p J((\mu,0), \vec{p}, - (\mu,0) - \vec{p})
\ee
where $J$ is defined in (\ref{defJ}), and $A$ and $B$ are the regions shown in Fig.~\ref{phi3region}A (symmetric between the vertical axis shown and the directions not depicted in the case $d>2$).  Here $\omega_{d} = 2\pi^{(d+1)/2}/\Gamma((d+1)/2)$ is the volume of the unit d-sphere.

\begin{figure}
\centering
\includegraphics[width=0.45\textwidth]{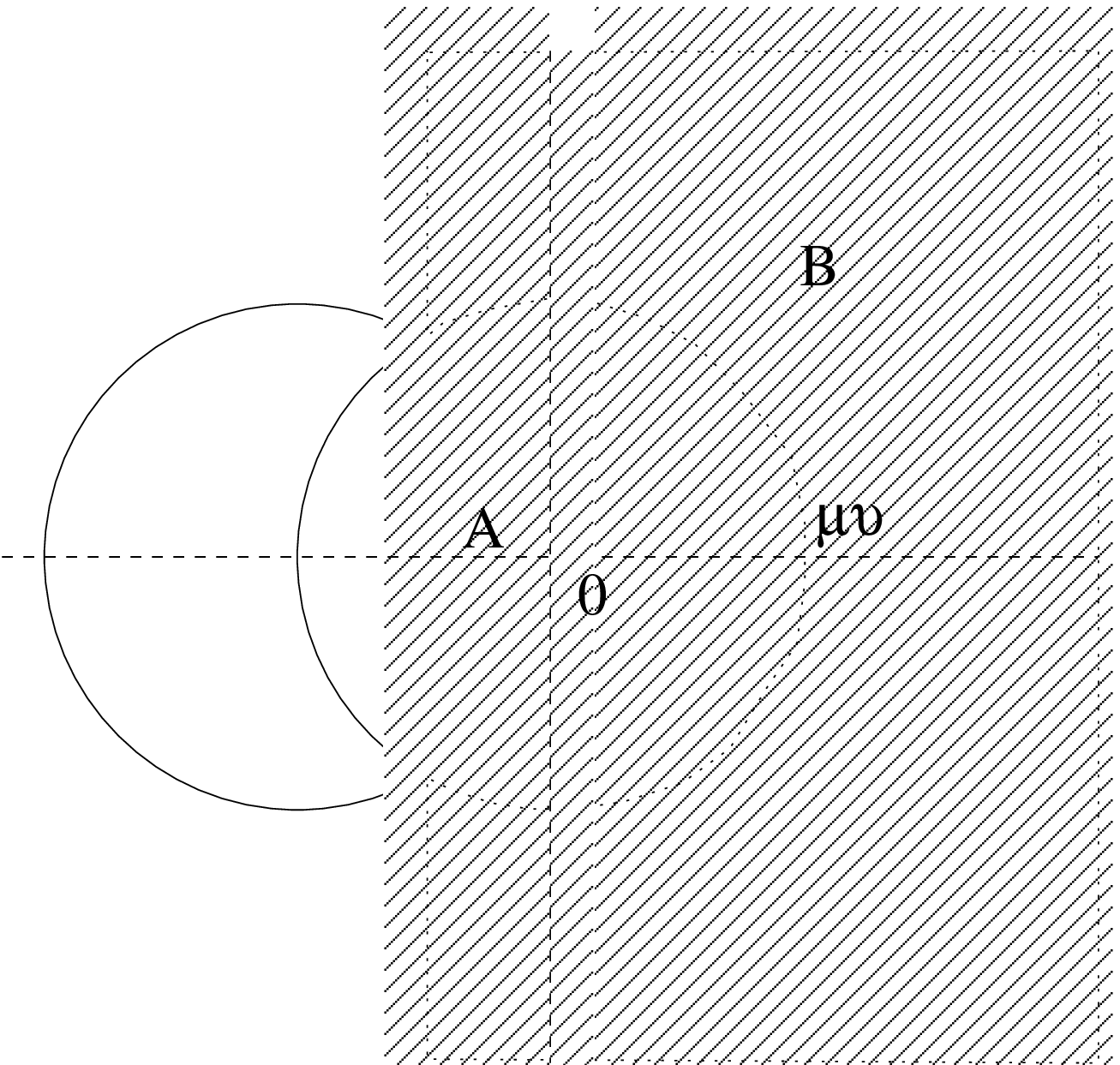}
\hfil
\includegraphics[width=0.45\textwidth]{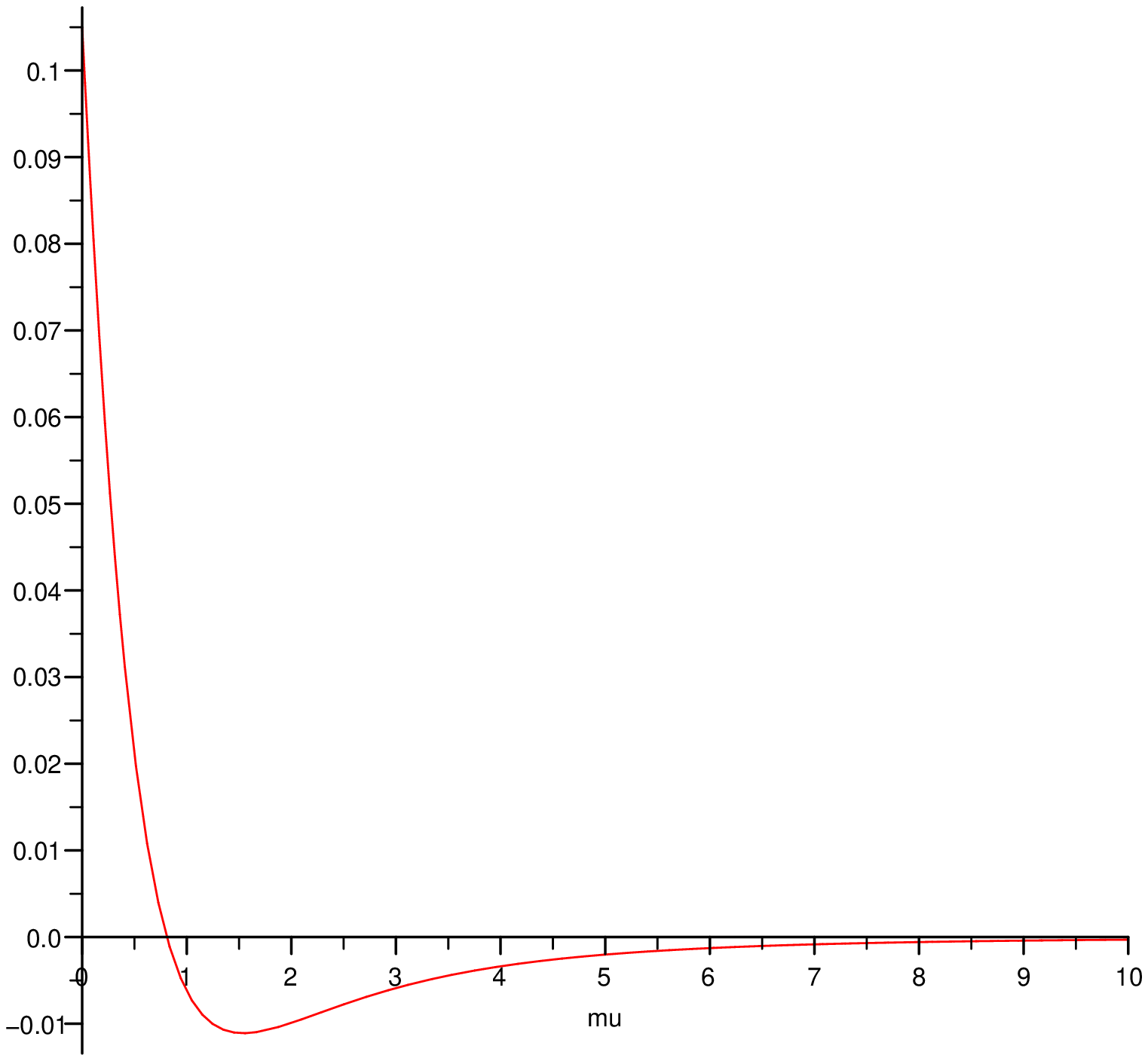} \\
({\bf A}) \hfil ({\bf B})
\caption{({\bf A}) Integration regions for $\phi^3$ theory in 2+1 dimensions.  ({\bf B}) The function $F(x)$ appearing in the entanglement entropy for $\phi^4$ theory in $1+1$ dimensions.}
\Label{phi3region}
\end{figure}

Explicitly, we have
\bea
{1 \over \omega_{d-1} \mu^{d-1}} {dI \over d \mu} &=& - \int^0_{-{\mu \over 2}} d p_x \int_0^{\sqrt{\mu^2 - (p_x^2 + \mu)^2}} d p_T \omega_{d-2} p_T^{d-2} J(p_x,p_T) \cr
&& + \int_{-{\mu \over 2}}^{\mu} d p_x \int_{\sqrt{\mu^2 - p_x^2}}^\infty d p_T \omega_{d-2} p_T^{d-2} J(p_x,p_T) \cr
&& + \int_{\mu}^{\infty} d p_x \int_0^\infty d p_T \omega_{d-2} p_T^{d-2} J(p_x,p_T)
\label{dIint}
\eea
where
\beas
J(p_x,p_T) &=& {1 \over \sqrt{\mu^2 + m^2} \sqrt{p_x^2 + p_T^2 + m^2} \sqrt{(\mu + p_x)^2 + p_T^2 + m^2} } \cr
&& \qquad  \cdot {1 \over \sqrt{\mu^2 + m^2} + \sqrt{p_x^2 + p_T^2 + m^2} + \sqrt{(\mu + p_x)^2 + p_T^2 + m^2}} \; .
\eeas
We find that
in 2 + 1 dimensions, the entanglement entropy decreases with $\mu$ as
\be
I_2(\mu)  \to  {2 \pi \over 3 \mu}
\ee
when $\mu \gg m$, while in 3+1 dimensions, we have
\[
I_3(\mu)  \to  8 \pi^2 (1-{1 \over 2} \ln(2)) \mu
\]
for $\mu \gg m$.  We interpret the the $3+1$ dimensional result as saying that in this case  the $\mu^3$ growth in the number of degrees of freedom below scale $\mu$ overwhelms the $1/\mu$ falloff of the effective dimensionless coupling.   These  expressions are exact (and  finite) as $m \to 0$.   For 4+1 dimensions and higher,  (\ref{dIint}) diverges -- we will discuss this divergence below.

\subsection{$\phi^4$ theory}

Finally, consider the $\phi^4$ field theory in 1+1 dimensions. From (\ref{result2}),
\beas
S(\mu)/V &=&  - \lambda^2 \log(\lambda^2) { 1 \over 16 (2 \pi)^{3}} \int_{ \{p_i\}_\mu} \prod d^d p_i {\delta(p_1 + \dots + p_4) \over  \omega_1 \cdots \omega_4 (\omega_1 + \dots + \omega_4)^2 } + {\cal O} (\lambda^2) \cr
&\equiv& - \lambda^2 \log(\lambda^2) { 1 \over 16 (2 \pi)^{3}} I(\mu) \; .
\eeas
Thus, we study
$I(\mu) = \int_{ \{p_i\}_\mu} dp_1 dp_2 dp_3 dp_4 \, \delta(p_1 + p_2 + p_3 + p_4) \, J(p_1,p_2,p_3,p_4)$,
where
$J(p_1,p_2,p_3,p_4)^{-1} = \omega_1 \omega_2 \omega_3 \omega_4 (\omega_1 + \omega_2 + \omega_3 + \omega_4)^2$.
It is again more convenient to evaluate
\beas
{1 \over 2}{d I \over d \mu} &=&
\left\{
\int_{\mu}^\infty d p_1 \int_{ \mu}^{p_1} d p_2
+ \int_{\mu}^\infty d p_1 \int_{ -{p_1+ \mu \over 2}}^{-\mu} d p_2 \right\} J(p_1,p_2,\mu,-p_1-p_2-\mu) \cr
 &-& \int_{-\mu}^{-{\mu \over 3}} d p_1 \int_{p_1}^{-{p_1 + \mu \over 2}} d p_2 \, J(p_1,p_2,\mu,-p_1-p_2-\mu)
\qquad \equiv  \qquad {1 \over m^4} F(\mu/m) \; .
\eeas
A numerical integration determines $F$, giving the final result
\be
S(\mu)/V = - \lambda^2 \log(\lambda^2) {1 \over 384 \pi^3} {1 \over m^3} \int_0^{\mu/m} dx F(x) \; .
\ee
The function $F(x)$ is plotted in Fig.~\ref{phi3region}B. By analyzing (analytically) the behavior of $F$ for large and small $x$, we find that the entropy $S(\mu)$ behaves as $\mu/m^4$ for small $\mu$ and as
\[
S \sim {1 \over \mu^3} \ln^2(\mu/m)
\]
for large $\mu$. As for the $\phi^3$ theory, the decay at large $\mu$ is related to the fact that the $\phi^4$ theory in 1+1 dimensions is free in the UV.


The leading perturbative contribution to the entanglement entropy $S(\mu)$ of $\phi^4$ theory can be similarly evaluated in 2+1 dimensions. The integrals there are more difficult to evaluate numerically, but are convergent. For 3+1 and higher dimensions, the integral expression for the leading contribution to $S(\mu)$ in the $\phi^4$ theory has a UV divergence, which we discuss further below.

\subsection{General remarks}
\Label{remarks}

\paragraph{Massless limits: } We found above that in two and higher space dimensions, the entanglement entropy $S(\mu)$ has a finite limit as we take the mass to zero.  However, in 1+1 dimensions, the results for both $\phi^3$ theory and $\phi^4$ theory diverge in the massless limit.  These divergences suggest that $S(\mu)$ is not an ``infrared-safe'' quantity for a massless scalar theory in 1+1 dimensions. However, the ratio $S(\mu)/S(\mu_0)$ has a finite limit if we hold $\mu$ and $\mu_0$ fixed as
we take $m$ to zero. The result is
\be
S(\mu)/S(\mu_0) = (\mu_0/\mu)^3
\ee
Thus, while it may not be sensible to talk about $S(\mu)$ directly for $m = 0$ and infinite volume, the ratio for different scales appears to be well-defined even in $1+1$ dimensions.

\paragraph{General understanding of large $\mu$ behavior: } The results above agree with the following heuristic derivation of the power law behavior of $S(\mu)$ for large $\mu$. The behavior is influenced by two significant effects. First, the number of degrees of freedom per unit volume below a momentum scale $\mu$ grows like $\mu^d$. All else being equal, we expect $S$ to scale like the number of degrees of freedom (for example, it is extensive). However the interactions in a general field theory depend on the scale, and this scale dependence also contributes to the behavior of $S(\mu)$. The dimensionless effective coupling for a $\phi^n$ interaction at scale $\mu$ behaves as $1/\mu^{d+1-n(d-1)/2}$. Since $S$ goes like $\lambda^2$ (plus logarithmic corrections), we can estimate that $S(\mu)$ should behave as
\[
S \sim \mu^d \times \left( {1 \over \mu^{d+1-n(d-1)/2}} \right)^2 = {1 \over \mu^{d+2 - n(d-1)}}
\]
up to possible logarithmic corrections. This is consistent with our results.  At a technical level this scaling arises in the integrals for entanglement entropy from two sources: (a) the measure factors (i.e. the density of modes), and (b) the energy denominators in the  interaction terms.   These are the same ingredients that affect the scaling of physical observables during renormalization.


\paragraph{Divergences: }
In various specific case considered above, we found that the leading perturbative contribution to $S(\mu)/V$ is finite in the limit where the IR and UV cutoffs are removed. However, for the $\phi^3$ theory in 4+1 and higher dimensions or the $\phi^4$ theory in 3+1 and higher dimensions, the integral expressions for the leading perturbative contribution to $S(\mu)$ diverge. The divergence is associated with the sum over states in (\ref{master}), or in the sum or integral over the momenta in (\ref{result}) or (\ref{result2}) respectively.  The leading divergence comes from the sum over states where one momentum has magnitude less than $\mu$ while the rest have magnitudes greater than $\mu$. The divergence is proportional to a power (or logarithm) of the UV cutoff $\Lambda$.

Of course, ultraviolet divergences are commonplace in quantum field theory. Typically, they are associated with integrals over momenta that appear beyond the leading order in perturbative calculations, and are dealt with by expressing the results in terms of renormalized (physical) parameters rather than bare parameters. However, here the divergences appear in leading order perturbative results. Since the bare and renormalized parameters are the same at leading order in perturbation theory, the divergences will not be eliminated by expressing the results in terms of renormalized parameters.\footnote{We do expect the standard divergences to appear in higher order perturbative corrections, even for quantities whose leading order result is finite. These divergences should be cured in the usual way by expressing results in terms of physical parameters, or by using renormalized perturbation theory with the appropriate counterterms.}

Such divergences in leading order expressions indicate a breakdown of perturbation theory for the specific quantity that is diverging. To see this, note first  that similar divergences appear even in fermionic theories, for example fermions in 1+1 dimensions with a $(\bar{\psi} \psi)^2$ interaction (see Appendix C for details). Furthermore, the divergence is present even at finite volume, since it is associated with the infinite number of high-momentum modes which are still present with an IR regulator. But for a theory of fermionic fields at finite volume, the Hilbert space associated with degrees of freedom with momentum below a scale $\mu$ is finite-dimensional. In this case, there is an upper bound $S(\mu) < \log (N)$, where $N$ is the dimension of the Hilbert space.   Now, consider the theory with a UV cutoff $\Lambda$. The leading perturbative expression for $S(\mu)$ will be finite for any finite $\Lambda$. But since this expression diverges as $\Lambda$ is taken to infinity, there will be some finite $\Lambda$ above which this leading contribution to $S(\mu)$ is larger than the bound $\log(N)$. Here, $\Lambda$ is still finite, so $S(\mu)$ is clearly well-defined, and the correct result for $S(\mu)$ must certainly be less than $\log(N)$, so the only possibility is that the leading perturbative expression is not a good approximation to the correct result.

Our conclusion should not be particulary surprising: regardless of how small the coupling parameter of a theory is, there will always be quantities that cannot be computed in perturbation theory. Here, the breakdown of perturbation theory seems to be associated with computing the entanglement entropy between a finite set of modes with the infinite set of degrees of freedom above scale $\mu$. We will see below that in cases where perturbation theory breaks down for this quantity, it is still possible to perturbatively calculate less inclusive quantities, such as the mutual information between degrees of freedom associated with two finite regions of momentum space. In cases where no divergence appears at leading order, the finite leading-order perturbative result should be reliable so long as subsequent terms in the perturbative expansion (after renormalization) are small compared to the leading terms.

\section{The extent of entanglement between scales}
\Label{single}

So far, we have considered the entanglement between modes in a field theory above and below some scale $\mu$. In this section, we ask about the entanglement entropy associated with a single mode of the field theory, or the mutual information between two individual modes. A version of the former observable has been considered previously in the condensed matter literature (see e.g. \cite{onemode}).   We can also consider the entanglement entropies of bounded  regions of momentum space. These sorts of observables are useful for two reasons: (a) they can be finite even when the entanglement entropy for the low-energy density matrix diverges, (b) they are a much more sensitive and clear probe of the extent of entanglement since they don't sum over the entire tower of UV modes.

\subsection{An aggregate measure of the range of entanglement}

The quantity $S(\mu)$ measures entanglement between the complete set of degrees of freedom below the scale $\mu$ and the complete set of degrees of freedom above the scale $\mu$.  Is this entanglement largely between modes just above and below the scale $\mu$, or is the entanglement  ``long-range'' in momentum space?

One way to address this question is to consider the entanglement entropy for the annular region $\mu_1 \le |p| \le \mu_2$ in momentum space. If the entanglement is  short-range, then for $\mu_2 \gg \mu_1$, the entanglement entropy $S([\mu_1,\mu_2]) \equiv S(\mu_1 \le |p| \le \mu_2)$ should be dominated by entanglement between modes just above and below the scales $\mu_1$ and $\mu_2$.   In addition, these separate contributions to the entanglement entropy should be well measured by $S(\mu_2)$ and $S(\mu_1)$.   Thus, for short-range entanglement we would expect
\be
S([\mu_1,\mu_2]) \approx S(\mu_1) + S(\mu_2) \qquad \qquad \mu_2 \gg \mu_1 \; .
\Label{shortrange1}
\ee
Alternatively, consider the mutual information between the degrees of freedom with $|p| \ge \mu_2$ and  $|p| \le \mu_1$:
\be
I(\mu_1,\mu_2) = S(\mu_1) + S(\mu_2) - S([\mu_1,\mu_2]) \, .
\Label{mut}
\ee
For short-range momentum space entanglement we expect (\ref{shortrange1}).  Hence, when $\mu_2 \gg \mu_1$ we expect that $I(\mu_1,\mu_2)  \approx 0$.   The rate of falloff of $I(\mu_2,\mu_1)$ as $\mu_2 / \mu_1$ increases from $1$ is a characterization of the extent of entanglement.

In $\phi^4$ theory the infinite volume expression for $I(\mu_1, \mu_2)$ is (using (\ref{master2}))
\be
\label{result2I}
S([\mu_1,\mu_2])/V =  - \lambda^2 \log(\lambda^2) {1 \over 24} \int_* \prod_i {d^d p_i \over 2 (2 \pi)^d} { (2 \pi)^d \delta(p_1 + p_2 + p_3 + p_4) \over (\omega_1 + \omega_2 + \omega_3 + \omega_4)^2 \omega_1 \omega_2 \omega_3 \omega_4} + {\cal O} (\lambda^2)
\ee
where the asterisk indicates that we integrate over momenta such that at least one $|p|$ is in the range $[\mu_1,\mu_2]$ and at least one $|p|$ is not in this range. For simplicity, we specialize to $d=1$ and take the mass $m=1$. It is simplest to first evaluate the quantity
\be
{d^2 S \over d \mu_1 d \mu_2} \; .
\ee
We can see that the only contribution to this will be from regions of the integral above where one momentum is at $\mu_1$ and another momentum is at $\mu_2$.

This is equal to the integral above with $p_1 = \pm \mu_2$, $p_2 = \pm \mu_1$, and the remaining $|p|$s either both inside or both outside the interval $[\mu_1,\mu_2]$. The distinct choices of momenta satisfying these constraints are
\beas
&p_1 = \mu_2 \qquad p_2 = \mu_1 \qquad p_3 \in (-\infty, -2 \mu_2 - \mu_1] \cup [-2 \mu_1 - \mu_2, -{1 \over 2}(\mu_1 + \mu_2)] \cr
&p_1 = \mu_2 \qquad p_2 = -\mu_1 \qquad p_3 \in (-\infty, -2 \mu_2 + \mu_1] \cup [- \mu_1 , {1 \over 2}(\mu_1 - \mu_2)] \cr
\eeas
or momenta obtained from these via $p_i \to -p_i$, where in all cases, $p_4$ is determined by the delta function constraint. Thus, we have
\beas
{1 \over  V} {d^2 S \over d \mu_1 d \mu_2} = - \lambda^2 \log(\lambda^2) {1 \over 12} {1 \over 16 (2 \pi)^3} \big\{ & \int_{-\infty}^{-2 \mu_2 - \mu_1} d p \; J(\mu_2,\mu_1,p,-p-\mu_1-\mu_2)  \cr
& + \int_{-2 \mu_1 - \mu_2}^{-{1 \over 2}(\mu_1 + \mu_2)} d p \; J(\mu_2,\mu_1,p,-p-\mu_1-\mu_2)  \cr
& + \int_{-\infty}^{-2 \mu_2 + \mu_1} d p \; J(\mu_2,-\mu_1,p,-p+\mu_1-\mu_2) \cr
& +   \int_{- \mu_1}^{{1 \over 2}(\mu_1 - \mu_2)} d p \; J(\mu_2,-\mu_1,p,-p+\mu_1-\mu_2) \big\}
\eeas
where
\[
J(p_1,p_2,p_3,p_4) = {1 \over (\omega_1 + \omega_2 + \omega_3 + \omega_4)^2 \omega_1 \omega_2 \omega_3 \omega_4} \; .
\]

To determine $S(\mu_1, \mu_2)$ from this expression, we can use $S(\mu, \mu) = 0$,  $S(0, \mu) = S(\mu)$, and ${\partial S \over \partial \mu_2} (0,\mu) = {d S \over d \mu}(\mu)$. From these, we have
\be
\label{der}
{\partial S \over \partial \mu_2}(\mu_1, \mu_2) = \int_0^{\mu_1} d\tilde{\mu}_1 {d^2 S \over d \mu_1 d \mu_2} (\tilde{\mu}_1, \mu_2) + {d S \over d \mu}(\mu_2)
\ee
and
\be
S(\mu_1, \mu_2) = \int_{\mu_1}^{\mu_2} d \tilde{\mu}_2 {d S \over d \mu_2}(\mu_1, \tilde{\mu}_2) = \int_{\mu_1}^{\mu_2} d \tilde{\mu}_2 \int_0^{\mu_1} d\tilde{\mu}_1 {d^2 S \over d \mu_1 d \mu_2} (\tilde{\mu}_1, \tilde{\mu}_2) + \int_{\mu_1}^{\mu_2} d \tilde{\mu}_2 {d S \over d \mu}(\tilde{\mu}_2) \; .
\ee
Here, $S(\mu)$ is the quantity that we evaluated in previous sections.

To investigate whether the entropy $S(\mu_1, \mu_2)$ is dominated by entanglement between degrees of freedom close to $\mu_1$ and $\mu_2$, we can vary $\mu_2$ and ask whether the variation of $S(\mu_1, \mu_2)$ is well approximated by the variation of $S(\mu_2)$ (these variations would be equal if $S(\mu_1,\mu_2) = S(\mu_1) + S(\mu_2)$). From (\ref{der}),  the difference between the variations is equal to the first term on the right hand side, so we ask whether this term is small compared with the other term. In Fig.~\ref{fig:entangfalloff}, we plot the ratio of these terms as a function of $\mu = (\mu_2 - \mu_1)/m$.  The ration declines  as $1/\ln\mu$ increases and approaches a finite value as $(\mu_2 - \mu_1)/m \to 0$.

\begin{figure}
\centering
\includegraphics[width=0.35\textwidth,angle=270]{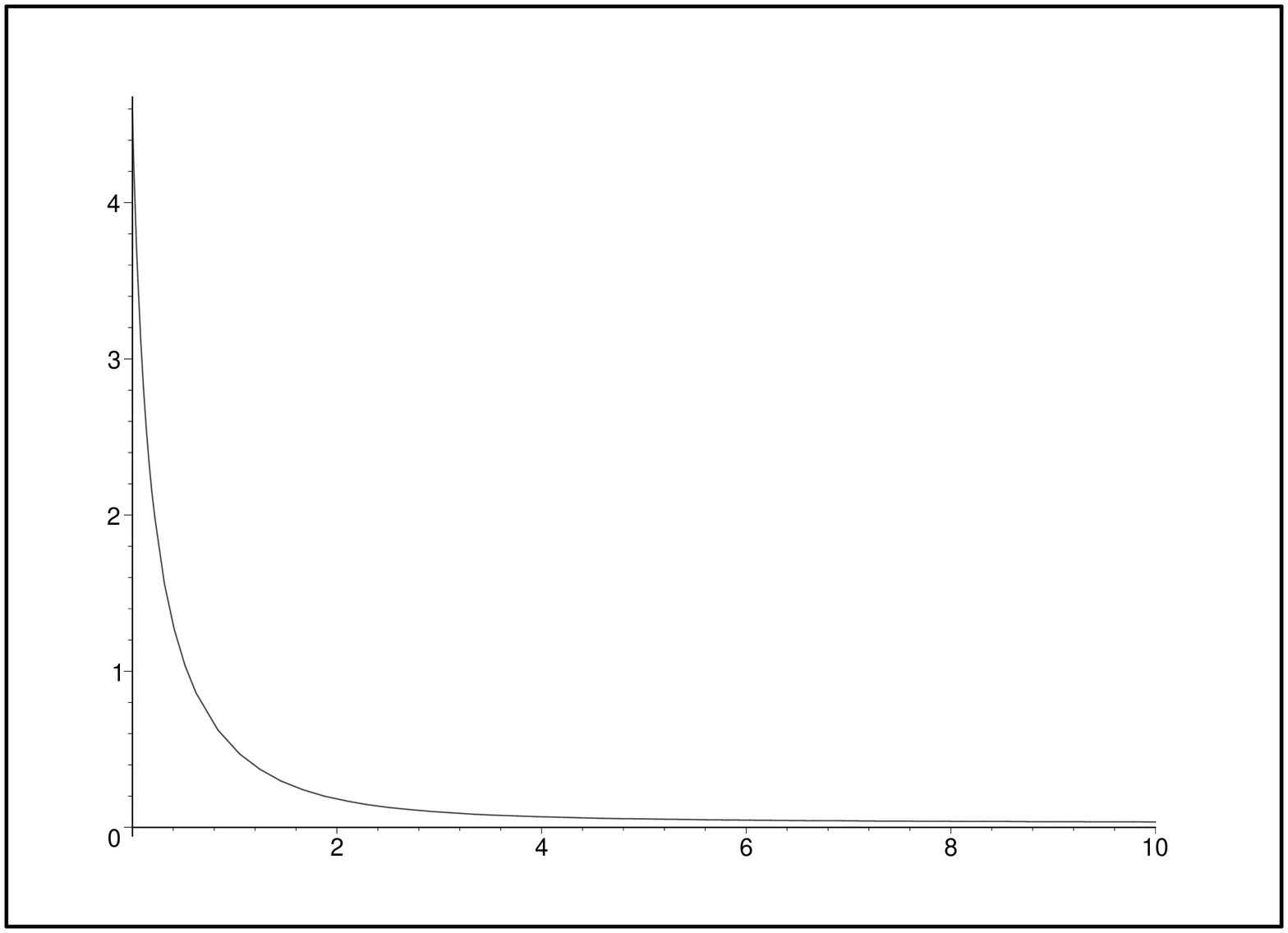}
\hfil
\includegraphics[width=0.35\textwidth,angle=270]{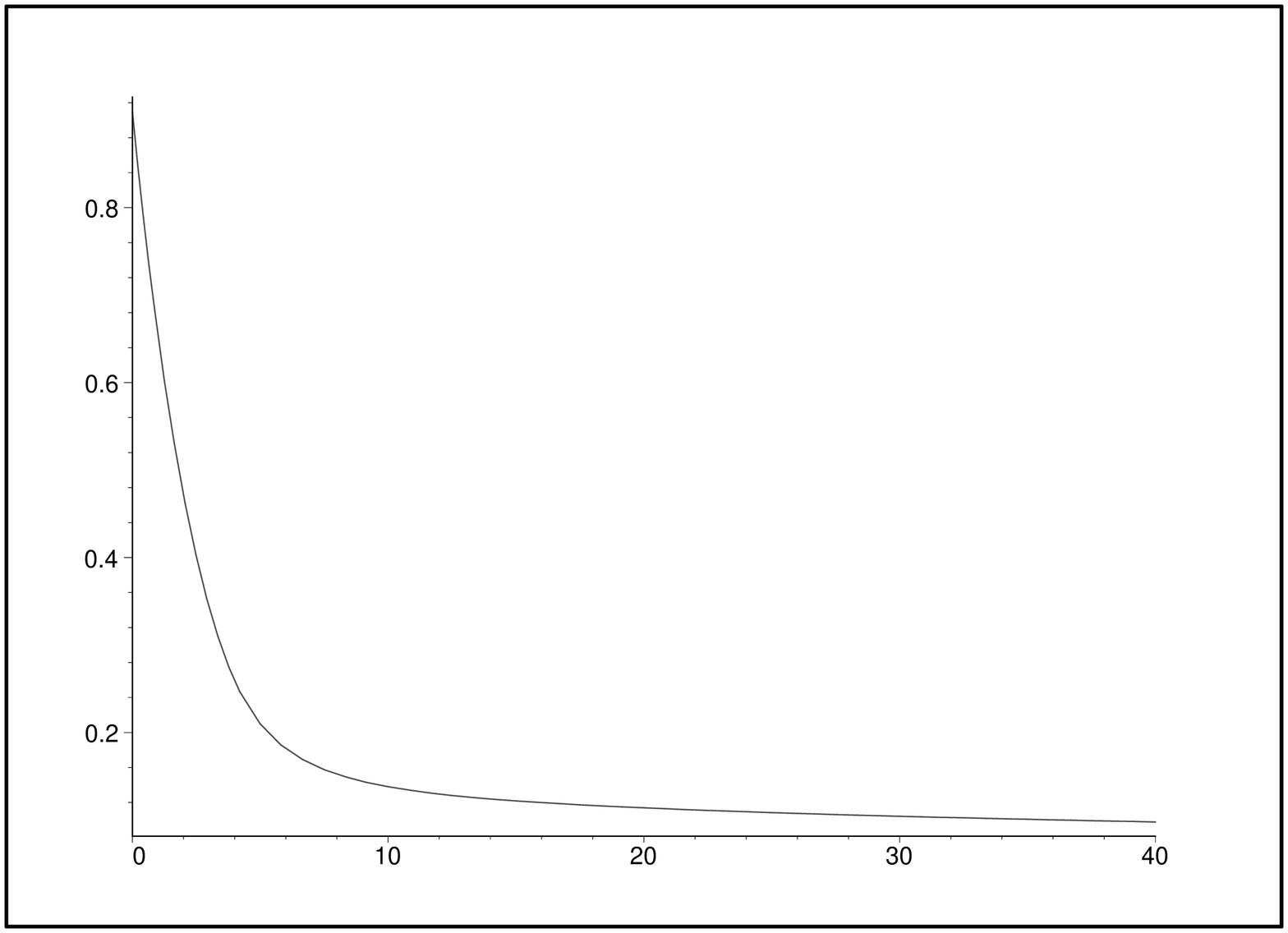} \\
({\bf A}) \hfil ({\bf B})
\caption{Ratio of first and second terms in (\ref{der}) vs $\mu  = (\mu_2 - \mu_1)/m$ for ({\bf A}) $\mu_1 = 1$ and ({\bf B}) $\mu_1 = 4$.  This is a measure of the range of entanglement in $\phi^4$ theory in $1+1$ dimensions. We have taken the mass to be $m=1$.}
\Label{fig:entangfalloff}
\end{figure}

The slow rate of decline is surprising given that the $\phi^4$ theory in 1+1 dimensions enjoys the property of decoupling.  Note however, that the quantity we are computing integrates over all of the UV modes.  Thus it is an aggregate measure of entanglement.     A more refined way to ask about the range of entanglement in momentum space is to consider the mutual information between individual modes at two different momenta $p$ and $q$ as we do below.   We will see that this mutual information falls off as a power law with $|q|$ when $|q| \gg |p|$.

\subsection{Single mode entanglement}

In this section, we calculate the entanglement entropy for a single mode with momentum $\vec{p}$. This measures the entanglement between a single mode and the rest of the field theory. The leading result for a $\phi^n$ scalar field theory follows immediately from (\ref{result}):
\be
S(\vec{p}) =  - \lambda^2 \log(\lambda^2) \sum_{ \{p_2,\dots,p_n\}} {\delta_{p + p_2 + \dots + p_n} \over 2^n L^{d(n-2)} \omega_p \omega_2 \cdots \omega_n (\omega_p + \dots + \omega_n)^2 } + {\cal O} (\lambda^2)
\ee
where the sum is over all distinct sets of $(n-1)$ momenta.\footnote{For $\vec{p}=0$, we have the further restriction that not all momenta are zero.} In the infinite volume limit, this gives
\beas
S(\vec{p}) &=&  - \lambda^2 \log(\lambda^2) {1 \over 2^n (2 \pi)^{d(n-2)}} \int_{ \{p_2,\dots,p_n\}} \prod_{i=2}^n d^d p_i {\delta^d(p + p_2 + \dots + p_n) \over  \omega_p \omega_2 \cdots \omega_n (\omega_p + \dots + \omega_n)^2 } + {\cal O} (\lambda^2) \cr
&\equiv& s_1(|\vec{p}|) \; .
\eeas
By rotational invariance, the result is a function only of $|p|$.  All explicit factors of the volume have canceled out without dividing by volume on the left side.

A natural interpretation of this {\it finite} quantity is that it gives the entanglement entropy density for degrees of freedom in an infinitesimal range $d^d p$ around the momentum $\vec{p}$. The number of modes in the box $d^d p$ is proportional to spatial volume, so if the entanglement entropy for one mode has no explicit volume dependence, the entanglement entropy for the set of modes in the box should be proportional to volume. This entropy is also proportional to the momentum space volume $d^d p$ (if this is infinitesimal), so the entanglement entropy associated with an infinitesimal volume $d^d x$ in position space and volume $d^d p$ in momentum space takes the form
\be
dS( \vec{p}) = {d^d x  \, d^d p \over (2 \pi)^d} \; s_1(|\vec{p}|) \; .
\ee
It is interesting that the phase space volume appears naturally here.\footnote{Note that while this entropy is spatially extensive, it is not extensive in momentum space. That it, it is not true that $S(R)/V = \int_R d^d p f(p)$.}

As an explicit example, we have plotted $s_1(p)$ for $\phi^3$ theory in two, three, and four spacetime dimensions in Fig.~\ref{fig:singmode}. In the figure, the entropies are normalized by their value at $p=0$. For $1+1$, $2+1$, and $3+1$ dimensions, the entanglement entropy decreases like $1/p^4$,$1/p^3$, and $1/p^2$ respectively.  Thus we see that in this case the entanglement of a {\it single mode} with the rest of the theory declines as power-law of the momentum, even though we found above that the {\it integrated} entanglement between modes above and below scales $\mu_2$ and $\mu_1$ only declines logarithmically.  The slow decay in the latter case is arising from the sum over modes.

\begin{figure}
\centering
\includegraphics[width=0.5\textwidth,angle=270]{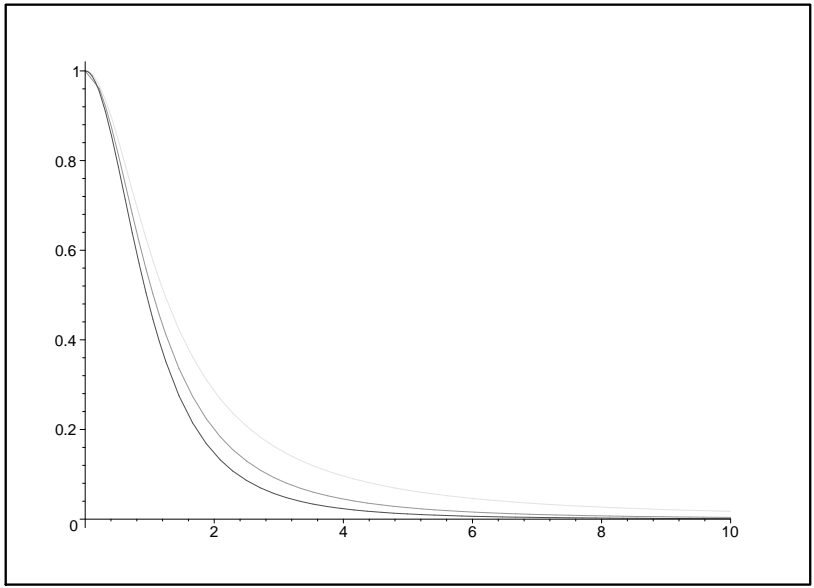}
\caption{Single-mode entanglement entropy vs magnitude of mode momentum for $\phi^3$ field theory in 1+1 (bottom), 2+1 (middle), and 3+1 (top) dimensions.  The entropies are normalized by their values at $p=0$.}
\Label{fig:singmode}
\end{figure}

\subsection{Mutual information between individual modes}

It is also interesting to investigate the mutual information between two specific field theory momenta. In the large volume limit, the natural quantity to consider is the mutual information between degrees of freedom in an infinitesimal range $d^d p$ around some momentum $p$ and degrees of freedom in an infinitesimal range $d^d q$ around some momentum $q$. Starting from the basic formula (\ref{master2}), only contributions from the second term in curly brackets survive the large volume limit. These involve matrix elements between the vacuum state and states where one particle is excited in each of the regions $d^d p$ and $d^d q$, and the remaining particles lie outside these regions. The resulting mutual information is proportional to $d^d p$ and $d^d q$ and to spatial volume, so we have:
\be
I(\vec{p},\vec{q})/V = d^d p \,  d^d q  \, {\cal I}(\vec{p},\vec{q}) ; .
\ee

For $\phi^n$ scalar field theory in $d+1$ dimensions, the leading contribution to ${\cal I}$ is
\be
\label{Igen}
{\cal I}(\vec{p},\vec{q}) = - \lambda^2 \log(\lambda^2) {1 \over 2^n (2 \pi)^{d(n-1)}} \int_{ \{p_3,\dots,p_n\}} \prod_{i=3}^n d^d p_i {\delta^d(p + q + p_3 + \dots + p_n) \over  \omega_p \omega_q \omega_3 \cdots \omega_n (\omega_p + \dots + \omega_n)^2 } + {\cal O} (\lambda^2)
\ee
where the integral is over distinct sets of $n-2$ momenta. For $\phi^3$ theory, this is
\be
{\cal I}(\vec{p},\vec{q}) = - \lambda^2 \log(\lambda^2) {1 \over 8 (2 \pi)^{2d}} {1 \over  \omega_p \omega_q \omega_{p+q}  (\omega_p + \omega_q + \omega_{p+q})^2 } + {\cal O} (\lambda^2)
\ee
Thus, the mutual information is enhanced when $\vec{p}$, $\vec{q}$ or $(\vec{p} + \vec{q})$ are near zero, and for fixed $p$, the mutual information falls off as $1/|q|^4$ for large $q$. While this expression gives the formal leading order result in any number of dimensions, we will see below that it should only be trusted as an accurate approximation to the exact result for $d \le 4$ space dimensions.

\subsection{Convergence and validity of leading order expressions.}

As for the entanglement entropy $S(\mu)$ considered in the previous section, the integrals in the leading order contributions to the  mutual information and entanglement entropy of single modes can contain UV divergences. As we argued in  Sec.~\ref{remarks}, such divergences indicate a breakdown of perturbation theory for the quantity in question.   In this subsection, we classify the scalar field theories for which the perturbative calculation of single-mode quantities $s_1(p)$ and ${\cal I}(p,q)$ gives sensible results.

We begin with the expression (\ref{Igen}) for the single mode mutual information of $\phi^n$ scalar field theory in $d+1$ spacetime dimensions. Naively, this will converge (i.e. there are enough powers of momenta being integrated over in the denominator) if
\be
d < 1 + {3 \over n-3} \; .
\ee
Thus, we have convergence in any dimension for $n=3$ for $d \le 3$ for $n=4$, for $d \le 2$ for $n=5$, and only for $d=1$ for any higher $n$.

Since higher order interactions (i.e. interactions with more powers of the field) are more likely to lead to divergences, we should be concerned that such higher order interactions generated in the quantum effective action will produce divergences at higher orders in perturbation theory. In $\phi^3$ theory we get an effective $\phi^n$ vertex at order $\lambda^n$ from a one loop diagram. As a function of the external momenta, this scales like $1/p^{2n-d+1}$ as these momenta are taken large. The contribution to $I(p,q)$ from such a vertex will naively be convergent if
\be
d < 5 + {3 \over n-1} \; .
\label{conv}
\ee
This is satisfied for any $n$ so long as $d \le 5$, but leads to a divergence in 6 and higher space dimensions. Thus, it appears that $I(p,q)$ can be computed in perturbation theory for $\phi^3$ theory in $d \le 5$ (the same dimensions for which the theory is renormalizible).

For $\phi^4$ theory, the effective action contains effective $\phi^{2n}$ interactions coming from one-loop diagrams at order $\lambda^n$. These scale with external momenta like $1/p^{2n-d+1}$. The contribution to $I(p,q)$ from such a vertex will naively be convergent if
\[
d < 3 + {1 \over 2 n -1} \; ,
\]
which is satisfied for any $n$ as long as $d \le 3$. Thus, it appears that $I(p,q)$ is a well defined quantity for $\phi^4$ theory in $d \le 3$ (again, the same dimensions for which the theory is renormalizible).

 An almost identical analysis shows that the mutual information between degrees of freedom in any two finite region of momentum space converges whenever $I(p,q)$ converges. Note that it would be incorrect to suppose from the considerations above that the leading order $I(p,q)$ is necessarily well defined for every renormalizable theory.  For example, according to (\ref{conv}), the leading order $I(p,q)$ diverges for the renormalizible $\phi^6$ theory in 3 dimensions.   We can also ask when the entanglement of a single mode  (or a finite region of momentum space) with the rest of the field theory is well defined. For $\phi^n$ theory, we find convergence for
\be
d < {2n-1 \over n-2}
\ee
This result extends to entanglement entropy of any finite region of momentum space.

\begin{table}[t]
\centering
\vskip 0.1 in
\begin{tabular}{| c |  c | c | }
\hline
Theory  & Dimensions where $I(A,B)$ converges & Dimensions where $S(A)$ converges \\   \hline   \, & &\\

$\phi^3$ & $d \le 6$ & $d \le 4$\\[0.1in]
$\phi^4$ & $d \le 4$ & $d \le 3$\\[0.1in]
$\phi^5$ & $d \le 3$ & $d =2$ \\[0.1in]
$\phi^{n \ge 6}$ & $d=2$ & $d=2$ \\[0.1in]
\hline
\end{tabular}
\caption{Dimensions where momentum space mutual information and entanglement entropy converge. The results apply for any bounded regions $A$ and $B$ in momentum space.}
\label{table:summary}
\end{table}

\section{Comments}

We have obtained a number of results for the scaling of entanglement entropies and mutual informations with the upper bound on a momentum interval. At a technical level, these results all follow from the density of modes (measure) in the integrals over momenta, and the energy denominators in the interactions.   These are the same ingredients that lead to the decoupling property of local quantum field theories.   Indeed, decoupling is usually understood simply as the power law suppression of higher dimension operators in a low energy effective theory.    This suppression means that high momentum degrees of freedom have weak effects on the dynamics of low-momentum degrees of freedom other than renormalizing the interaction strengths and wavefunctions.   Our study of entanglement between degrees of freedom with different momenta, and the resulting entanglement entropies and mutual informations,  refines this understanding of the influence between momentum scales.

In more detail, ``decoupling'' between UV and IR physics implies that starting from a generic action $S^\Lambda(g_I)$ at scale $\Lambda$, that  depends on an infinite number of parameters $g_I$, the Wilsonian effective action at a much lower scale $\mu$ will be very close to some action $S_W^\mu(g_i)$ in a family parameterized by a small number of physical parameters $g_i$.  In other words, the operation of integrating out degrees of freedom to successively lower scales results in a flow in the space of Wilsonian effective actions that converges to a low dimensional subspace at  scales  $\mu \ll \Lambda$.
  Now, according to (\ref{final}), the Wilsonian effective action $S^\mu_W$ at scale $\mu$ completely determines the reduced density matrix $\rho(\mu)$ for the degrees of freedom with $|p|< \mu$.   Thus we conclude that for the ground state of a generic field theory defined at scale $\Lambda$, the reduced density matrix for the degrees of freedom below some much lower scale $\mu$ will be very close to some family of density matrices $\rho(\mu,g_i)$ that depend on a small number of physical parameters $g_i$. Consequently, knowing the state of the low-momemtum degrees of freedom tells us relatively little about the details of the state at much higher scales.

The paucity of information about UV physics contained in the low-momentum density matrix should be reflected in some of the measures of quantum information we have discussed. Specifically, it seems likely that there is a connection between the decoupling behavior of field theories and the power-law fall off in mutual information observed in Sec.~\ref{single}.   It would be interesting to make this connection precise.



\paragraph{Relation to AdS/CFT: } In the context of gauge-theory / gravity duality (the AdS/CFT correspondence) \cite{malda}, there is now evidence that certain measures of entanglement in quantum field theory carry geometrical information about the dual spacetime. For field theories with a weakly curved dual gravity description, Ryu and Takayanagi  have proposed \red{rt} that the entanglement entropy for a spatial region $A$ is proportional to the area of the minimal surface $\tilde{A}$ in the bulk space whose boundary coincides with the boundary of $A$,
\[
S(A) = {Area(\tilde{A}) \over 4 G_N} \; .
\]
While the proposal has not yet been proven, it has passed a variety of checks (see e.g. \cite{Ryu:2006ef,Nishioka:2009un,Headrick:2010zt,Hayden:2011ag}).


Given the holographic interpretation of position-space entanglement entropy, it is natural to ask whether the momentum-space quantities considered in this paper have some simple dual geometrical interpretation for field theories with gravity duals. As an example, the quantity $S(\mu)$ measures the entanglement between degrees of freedom above and below the scale $\mu$. Since energy/momentum scale in holographic field theories corresponds to radial position in the dual geometry, we might guess that $S(\mu)/V$ is related to the area (per unit field theory volume) of a surface separating the IR region $r < r(\mu)$ of the dual geometry from the UV region $r > r(\mu)$. For the dual geometry to a translation-invariant field theory state, this area function is a well-defined observable.\footnote{If the spatial part of the dual metric is $dr^2 + f(r) dx^2$, the area of the surface at radius $r(\mu)$ per unit field theory volume is proportional to a power of $f(r(\mu))$.} However, we currently have no way to check whether this or some similar observable corresponds to momentum-space entanglement entropy, since we cannot calculate this entropy for any strongly coupled field theory with a gravity dual.\footnote{$S(\mu)$ will probably not always correspond in a simple way the specific area observable mentioned, since that area would be well-defined even for gravity duals of 0+1 dimensional field theories, for which there is no way to divide up the degrees of freedom by spatial momentum, and therefore no way to define $S(\mu)$.  Of course such low-dimensional gauge/gravity dualities  (e.g. AdS$_2$/CFT$_1$ also have many other special features \cite{Strominger:1998yg, Balasubramanian:2009bg}).}

\paragraph{Relation to DMRG and MERA:} Here we have explored various aspects of entanglement in quantum
field theory and the connection to renormalization theory. In the
condensed matter literature, the ideas of entanglement and renormalization
have come together previously in various schemes for approximating the
ground state of many-body systems \cite{DMRQ,MERA}. While the focus and details of that
work are rather different from the present discussion, it may be useful to
briefly review those ideas here.

Consider a quantum many-body system described by some lattice of degrees
of freedom, for which the Hilbert space decomposes as a tensor product of
Hilbert spaces for the individual sites. The dimension of the full Hilbert
space is $d^N$ where $d$ is the dimension of the individual Hilbert spaces
and $N$ is the number of sites. A general state (and in particular, the
exact ground state of the system for a given Hamiltonian) can be
represented exactly by a tensor $T^{i_1 \cdots i_N}$ that gives the
coefficient of the basis state $|i_1\rangle \otimes \cdots \otimes |i_N
\rangle$.

A general numerical determination of the ground state is impractical due
the the large number $d^N$ of independent coefficients. For certain
systems, usually in 1+1 dimensions, an efficient variational approach to
approximating the ground state is to consider tensors $T$ that can be
decomposed into contractions of lower-rank tensors. For example, the
``Matrix Product State'' (MPS) decomposition corresponds to
\[
T^{i_1 \cdots i_N} = (M_1)^{i_1}_{a_1 a_2} (M_2)^{i_2}_{a_2 a_3} \cdots
(M_N)^{i_N}_{a_N a_1} \; .
\]
In practice, one uses this decomposition as a variational ansatz, varying
the individual matrices $M^i$ to arrive at the best approximation to the
ground state \red{Cite}.  If the dimension of the matrices $M^i$ is large enough, any
tensor $T$ can be represented in this way, so the variational method gives
an exact result. However, for a wide class of systems, it has been found
that the ground state can be well approximated by matrices of much lower
dimension. In this case, the matrix product ansatz represents a truncation
of the Hilbert space to a subspace of lower dimension, and in cases where
it is effective, the true ground state is close to the ground state in
this subspace.

It turns out that the success of this method is related to the
entanglement properties of the ground state. The optimal method of
truncating to a lower-dimensional Hilbert space is to retain as much of
the entanglement entropy for the various subsystems (blocks of sites) as
possible.\footnote{This idea arose first in the ``Density Matrix
Renormalization Group" (DMRG)  \cite{DMRG} an iterative renormalization procedure on
the {\it state} of the system that truncates the Hilbert space in each
step while retaining as much entanglement entropy as possible. The DMRG is
now understood to give results equivalent to this MPS variational method.
} The procedure works most efficiently (i.e. for smallest matrices $M$)
when there is limited entanglement between the subsystems corresponding to
blocks of sites. For systems with a highly entangled ground state, the
method is much less efficient.

Another approach that is more successful in cases with long-range
entanglement is the ``Multiscale Entanglement Renormalization Ansatz''
(MERA \cite{MERA}). In this approach the tensor $T$ is represented by an iterative
procedure. The tensor is first written in terms of a ``disentangled''
tensor $\tilde{T}$ using unitary matrices $U$:
\[
T_{(n)}^{i_1 \cdots i_{2N}} = (U_1^{(n)})^{i_1 i_2}_{j_1 j_2} \cdots
(U_N^{(n)})^{i_{2N-1} i_{2N}}_{j_{2N-1} j_{2N}} \tilde{T}_{(n)}^{j_1
\cdots j_{2N}}
\]
and then $\tilde{T}$ is represented in terms of a lower rank tensor using
``projectors'' $P$:
\[
\tilde{T}_{(n)}^{j_1 \cdots j_{2N}} = (P_1^{(n)})^{j_1 j_2}_{I_1} \cdots
(P_N^{(n)})^{j_{2N-1} j_{2N}}_{I_N} T_{(n+1)}^{I_1 \cdots I_N} \; . \]
The latter step can be understood as a ``coarse-graining'' of the system,
though the dimension of the index space $I$ is not necessarily the same as
that of the original index space $i$. The original tensor $T_{(1)}$ is
thus represented by the individual matrices $(U_i^{(n)})^{i j}_{i' j'}$
and $(P_i^{(n)})^{ij}_{I}$ which are the variational parameters used to
approximate the ground state.

The introduction of the $U$ matrices is motivated by the observation that
coarse graining works most efficiently when there is little entanglement
between the adjacent blocks. The unitary matrices $U$ can remove
short-range entanglement between adjacent blocks before coarse graining.
In this way, the matrices $U_{(n)}$ encode the entanglement between sites
at the $n$th level, which corresponds in the original picture to blocks of
$2^n$ sites. Thus, in the MERA representation of a ground state the
unitary matrices $U$ encode entanglement at different scales. This
information is certainly related to the scale-dependent entanglement
entropies considered in this paper, though the MERA entanglements would
seem to be more closely related to position space entanglement. Also, the
original MERA applies only to discrete systems, though an extension to
continuum quantum field theories has been recently proposed in \cite{Haegeman:2011uy}. 
An interesting connection between  MERA and the AdS/CFT proposal above has been given in \cite{swingle}.

\paragraph{Acknowledgments: } We would like to thank Horacio Casini, Patrick Hayden, Matthew Headrick, Ting Chen Leo Hsu, Rob Myers, Hirosi Ooguri, John Preskill, Robert Raussendorf and Moshe Rozali for helpful discussions. VB is grateful to the theory group at UBC for hospitality while this work was initiated, and thanks the Aspen Center for Physics and the Santa Fe Insitute for hospitality while this work was completed.     VB is supported in part by DOE grant DOE grant DE-FG02-95ER40893.

\appendix

\section{Entanglement entropy at $O(\lambda^3)$}
\Label{app:higher}

The calculations in Sec.~\ref{sec:entpert} can be extended to give an expression for the ${\cal O}(\lambda^3)$ terms in the perturbative calculation of entanglement entropy for a general system. We find that after a similarity transformation, the density matrix  (\ref{densitymat1}) can be written as
\be
\hat{\rho}_A = \left( \ba{cc} 1 - |C|^2  + A^\dagger C B^\dagger + B C^\dagger A & 0 \cr 0 & C C^\dagger - A B C^\dagger - C B^\dagger A^\dagger \ea \right) + {\cal O}(\lambda^4) \, .
\ee
The eigenvalues of this matrix (up to corrections of order $\lambda^4$) include the top-left element of the matrix, and the eigenvalues of the lower-right matrix. At leading order, the lower-right matrix is $C_1 C^\dagger_1$ where $C_1$ is the order $\lambda$ term in $C$. We defined the eigenvalues of this matrix to be $a_i$. Finding the eigenvalues of the lower-right matrix after the higher order terms are added is a problem formally equivalent to ordinary time-independent perturbation theory in quantum mechanics, so we can express the result in terms of the eigenvalues and eigenvectors of $C_1 C^\dagger_1$.

Using this approach, the result for the entanglement entropy up to order $\lambda^3$ is
\beas
S(\mu) &=& \lambda^2 (-\log(\lambda^2) + 1) \tr(C_1 C_1^\dagger) - \lambda^2 \sum_i a_i \log(a_i)   \cr
&& \qquad + \lambda^3 (-\log(\lambda^2)) \tr(C_1 C_2^\dagger + C_2 C_1^\dagger- A_1 B_1 C_1^\dagger - C_1 B_1^\dagger A_1^\dagger)  \cr
&& \qquad - \lambda^3 \sum_i \log(a_i) \langle v_i | C_1 C_2^\dagger + C_2 C_1^\dagger- A_1 B_1 C_1^\dagger - C_1 B_1^\dagger A_1^\dagger | v_i \rangle
\eeas
where $a_i$ and $v_i$ are the eigenvalues and eigenvectors of the matrix $C_1 C_1^\dagger$ and the subscripts indicate the order in perturbation theory.

\section{Momentum-space entanglement and correlators}

Starting with the general expression (\ref{master}) for the leading order perturbative contribution to entanglement entropy, we can now specialize to the case of quantum field theory. We find that
\beas
S(P) &=& - \lambda^2 \log(\lambda^2) \sum_{n \ne 0, N \ne 0} { |\langle n,N | H_{AB} | 0, 0 \rangle|^2  \over (E_0 + \tilde{E}_0 - E_n - \tilde{E}_N)^2} + {\cal O} (\lambda^2) \cr
&=& - \lambda^2 \log(\lambda^2)  \sum_{n \ne 0, N \ne 0} \int_0^\infty d \tau \tau \langle  0, 0 | H_{I} |  n,N \rangle e^{(E_{0,0}-E_{n,N}) \tau}  \langle n,N | H_{I} | 0, 0 \rangle  + {\cal O} (\lambda^2)  \cr
&=& - \lambda^2 \log(\lambda^2) \sum_{n \ne 0, N \ne 0} \int_0^\infty d \tau \tau \langle  0, 0 | e^{H_0 \tau} H_{I} e^{-H_0 \tau} |  n,N \rangle   \langle n,N | H_{I} | 0, 0 \rangle  + {\cal O} (\lambda^2) \cr
&=& - \lambda^2 \log(\lambda^2) \int_0^\infty d \tau \tau \langle  0, 0 |e^{H_0 \tau} H_{I} e^{-H_0 \tau} \Pi_A H_{I} | 0, 0 \rangle  + {\cal O} (\lambda^2) \cr
&=& - \lambda^2 \log(\lambda^2) \int_0^\infty d \tau \tau \langle H_{I}(-i\tau) \Pi_A H_{I}(0) \rangle  + {\cal O} (\lambda^2) \cr
&=& - \lambda^2 \log(\lambda^2) \int_0^\infty d \tau \tau \int d^3 x d^3 y \langle {\cal H}_{I}(-i\tau,x) \Pi_A {\cal H}_{I}(0,y) \rangle  + {\cal O} (\lambda^2) \cr
&=& - V \lambda^2 \log(\lambda^2) \int_0^\infty d \tau \tau \int d^3 x \langle {\cal H}_{I}(-i\tau,x) \Pi_A {\cal H}_{I}(0,0) \rangle  + {\cal O} (\lambda^2)
\eeas
Here, we use the standard definition of time-dependent operators in the ``interaction picture'':
\[
H_I(t) \equiv e^{i H_0 t} H_{I} e^{-i H_0 t} \; .
\]
The operator $\Pi$ is projects to intermediate states with at least one particle having momentum in the subset $P$ and at least one particle having momentum in the complementary subset of momenta.

The factor of volume in the last line comes from the $y$ integral in the previous line, which is trivial since the correlator in that line can depend only on the combination $x-y$.  The  entropy per unit volume $S(P)/V$ will have a finite limit, so that $S(P)$ is an extensive quantity.

 \section{Entanglement entropy in a fermionic system}

 Here we calculate the entanglement entropy in a fermionic theory with a $(\bar{\psi} \psi)^2$ interaction.  Consider for definiteness the renormalizable theory in 1+1 dimensions.  The fermion fields are expanded as
 \begin{equation}
 \psi(x) =  \sum_p \frac{1}{L^{\frac{1}{2}}} \frac{1}{\sqrt{2 \omega_p}}  \(a_p u(p) e^{-ipx} + b^{\dagger}_p v(p) e^{ipx} \).
 \label{exp}
 \end{equation}
 As a straightforward application of (\ref{master}) the entanglement entropy is
\begin{equation}
S_{\mu} = -\lambda^2 \log(\lambda^2) \sum_t \sum_{p}^*  \frac{\left| \langle \{p,t\}_1, ... , \{p,t\}_4 \left|  \( \bar{\psi} \psi \)^2 \right| 0 \rangle \right|^2} { \( \omega_1 +\omega_2+\omega_3+\omega_4 \)^2} + O(\lambda^2),
\label{ent}
\end{equation}
 where t indicates the type of fermion (i.e. particle or antiparticle).  The star indicates that the sum over momenta is restricted to the set where at least one momentum is above and at least one momentum is below the scale $\mu$.
 Substituting the expansion (\ref{exp}) into (\ref{ent})
 \begin{equation}
S_{\mu} = -\lambda^2 \log(\lambda^2) \frac{6\cdot4}{2^{4} L^2} \sum_p^* \frac{ \delta_{\sum_i p_i}}{\(\sum_i \omega_{p_i} \)^2 \prod_i \omega_{p_i}}  \left|   \bar{u}(p_1) v(p_2) \bar{u}(p_3) v(p_4) - \bar{u}(p_1) v(p_4) \bar{u}(p_3) v(p_2)        \right|^2,
 \end{equation}
 where 6 is the number of ways or choosing 2 particles and 2 antiparticles.  Using 1+1 dimensional spinor and gamma matrix identities, and passing to the infinite volume limit we are left with
 \begin{equation*}
S_{\mu}/L =-\lambda^2 \log(\lambda^2) \frac{6}{(2 \pi)^{3}} \int^* dp_1 ... dp_4  \delta({\sum_i p_i) } \frac{(p_1 \cdot p_3 - m^2)(p_2 \cdot p_4 - m^2)}{\(\sum_i \omega_{p_i} \)^2 \prod_i \omega_{p_i}}.
 \end{equation*}
 In the region where three momenta are taken to be large this integral diverges linearly.

\end{document}